\documentclass[12pt]{article}

\usepackage[inline,shortlabels]{enumitem}
\usepackage{float}
\usepackage{mathtools}
\usepackage{graphicx}
\usepackage[round]{natbib}
\usepackage[margin=1in]{geometry}
\usepackage{tikz}
\usepackage[english]{babel}
\usepackage{longtable}
\usepackage{color}
\usepackage{amssymb,amsmath,amsthm}
\usepackage{multirow}
\usepackage[titletoc,title]{appendix}
\usepackage{authblk}
\usepackage{setspace}
\usepackage{dsfont}
\usepackage[OT1]{fontenc}
\usepackage{refcount}
\graphicspath{ {./figs/} }

\usepackage{nameref,zref-xr} 
\zxrsetup{toltxlabel} 

\usepackage{subfig}
\usepackage[inline]{trackchanges}
\addeditor{Ivan}
\addeditor{Nima}
\addeditor{Kara}

\newtheorem{theorem}{Theorem}
\AtEndDocument{\refstepcounter{theorem}\label{finalthm}}
\AtEndDocument{\refstepcounter{equation}\label{finaleq}}

{
  \theoremstyle{definition}
  
}
{
  \theoremstyle{definition}
  \newtheorem{assumptioniden}{}
}
{
  \theoremstyle{definition}
  
}

\newtheorem{lemma}{Lemma}
\AtEndDocument{\refstepcounter{lemma}\label{finallemma}}

\DeclareMathOperator{\expit}{expit}

\renewcommand{\P}{\mathsf{P}}

\newcommand{\cc}{\mathsf{c}}
\newcommand{\p}{\mathsf{p}}
\newcommand{\q}{\mathsf{q}}
\newcommand{\g}{\mathsf{g}}
\newcommand{\e}{\mathsf{e}}
\newcommand{\uu}{\mathsf{u}}
\newcommand{\h}{\mathsf{h}}
\newcommand{\vv}{\mathsf{v}}
\newcommand{\rr}{\mathsf{r}}

\newcommand{\indep}{\mbox{$\perp\!\!\!\perp$}} 
 \newcommand{\dd}{\mathrm{d}}

\newcommand{\one}{\mathds{1}}
 
\newcommand{\E}{\mathsf{E}}

\renewenvironment{proof}{{\it Proof }}{\qed \\}

\DeclarePairedDelimiterX{\norm}[1]{\lVert}{\rVert}{#1}

\pgfdeclarelayer{background}
\pgfsetlayers{background,main}
\usetikzlibrary{arrows,positioning}
\tikzset{
>=stealth',
punkt/.style={
rectangle,
rounded corners,
draw=black, very thick,
text width=6.5em,
minimum height=2em,
text centered},
pil/.style={
->,
thick,
shorten <=2pt,
shorten >=2pt,}
}
\newcommand{\Vertex}[2]
{\node[minimum width=0.6cm,inner sep=0.05cm] (#2) at (#1) {$\footnotesize#2$};
}
\newcommand{\Vertexr}[2]
{\node[rectangle, draw, minimum width=0.6cm,inner sep=0.05cm] (#2) at (#1) {$\footnotesize#2$};
}
\newcommand{\ArrowR}[3]%
{ \begin{pgfonlayer}{background}
\draw[->,#3] (#1) to[bend right=30] (#2);
\end{pgfonlayer}
}
\newcommand{\ArrowL}[3]%
{ \begin{pgfonlayer}{background}
\draw[->,#3] (#1) to[bend left=45] (#2);
\end{pgfonlayer}
}
\newcommand{\EdgeL}[3]%
{ \begin{pgfonlayer}{background}
\draw[dashed,#3] (#1) to[bend right=-45] (#2);
\end{pgfonlayer}
}

\newcommand{\Arrow}[3]%
{ \begin{pgfonlayer}{background}
\draw[->,#3] (#1) -- +(#2);
\end{pgfonlayer}
}
\newcommand{\titlepaper}{Causal mediation with instrumental variables}

\date{\today}
\author[1]{Kara E. Rudolph\thanks{corresponding author:\\
kr2854@cumc.columbia.edu \\ tel. +12123422926 }}
\author[1]{Nicholas Williams}
\author[2]{Iv\'an D\'iaz}

\affil[1]{\small Department of
  Epidemiology, Mailman School of Public Health, Columbia University.}
\affil[2]{\small Division of Biostatistics, Weill Cornell Medicine.}

\usepackage[colorlinks,citecolor=blue,urlcolor=blue]{hyperref}

\title{\titlepaper}

\begin{document}
\maketitle

\begin{abstract}
Mediation analysis is a strategy for understanding the mechanisms by which treatments or interventions affect later outcomes. Mediation analysis is frequently applied in randomized trial settings, but
typically assumes: a) that randomized assignment is the exposure of interest as opposed to actual take-up of the intervention, and b) no unobserved confounding of the mediator-outcome relationship.
In contrast to the rich literature on instrumental variable (IV) methods to estimate a total effect of a non-randomized exposure, there has been almost no research into using IV as an identification strategy in the presence of both exposure-outcome and mediator-outcome unobserved confounding.
In response, we define and identify novel estimands---complier interventional direct and indirect effects (i.e., IV mediational effects) in two scenarios: 1) with a single IV for the exposure, and 2) with two IVs, one for the exposure and another for the mediator, that may be related. We propose nonparametric, robust, efficient estimators, and apply them to a housing voucher experiment.
\end{abstract}

\noindent%
\vfill

\section{Introduction}
Mediation analysis is a strategy for understanding the mechanisms by which treatments or interventions affect later outcomes. Such treatments or interventions may be randomly assigned, as in randomized controlled trials. Mediation has been examined in many different types of trial settings; e.g., in examining the extent to which a randomized behavioral intervention works to reduce adolescent suicide behavior through improving family functioning \citep{pineda2013family}, quantifying the extent to randomized reductions in air pollution via air purifiers affects cardiovascular biomarkers through DNA hypomethylation \citep{chen2016dna}, and understanding how receiving a low-income housing voucher affects later adolescent mental health outcomes by operating through the changes in the school and neighborhood environments \citep{rudolph2018mediation}. Please see \citet{vo2020conduct} for a review of mediation analyses in the context of randomized trials. 

We use one of the above settings as the motivating example throughout this paper. Specifically, we consider a mediation scenario from the Moving to Opportunity (MTO) study, which was a large-scale trial in which families were randomized to receive a Section 8 housing voucher (the most common form of federal housing assistance in the United States) that they could use to move out of public housing and, instead, rent on the private market \citep{kling2007experimental,sanbonmatsu2011moving}. We are interested in the effect of moving with the voucher on 
risk of developing a psychiatric mood disorder during adolescence, possibly operating through the mediator of neighborhood poverty.

Current methods for causal mediation analyses in trial settings are generally limited in at least two aspects. First, current methods typically assume that randomized assignment is the exposure of interest as opposed to actual take-up of the treatment/intervention. Second, they rely on a no-unobserved confounders assumption in terms of the relationship between the mediators and outcome.

Current methods typically consider the following observed data scenario: a vector of random variables $O=(W,A,M,Y)$ is observed for $n$ individuals, where $W$ represents covariates, $A$ represents randomized assignment, $M$ represents mediators, and $Y$ represents the outcome. In the absence of perfect compliance, the effects of randomized treatment assignment (the intent-to-treat effects) may be different from the effects of the treatment actually taken (the as-treated effects) \citep{hernan2006instruments}. For example, in MTO motivating example, 
noncompliance was relatively common, with between 18-32\% of families who received a housing voucher not using it to move, depending on the city (e.g., because of difficulty finding landlords willing to accept the vouchers and other logistical challenges.)

Thus, in many cases, like the MTO exampled described above, observed data is more accurately represented by $O=(W,A,Z,M,Y),$ where $Z$ represents treatment actually taken, which is frequently the exposure of scientific interest. This motivates use of instrumental variable (IV) methods. In the case of non-mediated total effects, one can use an instrumental variable $A$ (which needs to adhere to several assumptions---e.g., randomized conditional on observables, monotonicity, and the exclusion restriction) to identify the effect of the treatment actually taken $Z$ (not randomly assigned) on an outcome $Y$, where the effect otherwise would not be identifiable due to unobserved confounding \citep{Angristetal&Imbens&Rubin96}. Such effects are typically called complier average causal effects \citep{imbens1997estimating}. The literature on identifying causal estimands with instrumental variables and estimating such parameters is vast \citep[e.g.,][]{angrist1995identification,abadie2003semiparametric,aronow2013beyond,chernozhukov2020instrumental}.

In contrast to the rich literature on instrumental variable methods in the case of total effects, there has been almost no development of IV methods for mediational direct and indirect effects. Previous work by \citet{rudolph2020complier} identified and estimated IV mediational effects in the observed data setting $O=(W,A,Z,M,Y).$ However, this work was limited in terms of only identifying and estimating the complier direct effect, assuming a known stochastic intervention on the mediator, and considering only a single instrument for the exposure, therefore still requiring the assumption of no unobserved confounding of the mediator-outcome relationship.

The limitation of needing to assume no unobserved confounding of the mediator-outcome relationship undercuts the main motivation for using a randomized design/IVs in the first place, which is to identify the effect of the exposure on the outcome ($Z$ on $Y$) in the presence of unobserved confounding, and if unobserved confounding is a concern for the $Z-Y$ relationship, it is likely also a concern for the $M-Y$ relationship. 
 To our knowledge, there has been almost no research into using IV as an identification strategy for both $Z$ and $M$, the exception being the work of \citet{frolich2017direct}. In this work, the authors enumerated identification strategies in such a mediation scenario, all of which necessitate an additional instrument---one instrument for $Z$ and another for $M$. The observed data can then be written $O=(W, A, Z, L, M, Y)$, where $A$ is the instrument for $Z$ and where $L$ is the instrument for $M$. \citet{frolich2017direct} proposed an estimator for a case where the instruments $A$ and $L$ are independent from one another, which may occur only rarely in practice. In addition the authors also assume
that the mediator is a function of this $L, A, W$, and exogenous, unobserved factors, $U$, where $U$ is a continuous variable with
strictly increasing cumulative distribution function, and where this function is strictly monotonic in $U$ \citep{frolich2017direct}.

To summarize, we know of no approach to estimate complier direct and indirect effects in the following general settings: 1) in the case of a single instrument for the exposure but none for the mediator, and 2) in the case of two, non-independent instruments, one for the exposure and one for the mediator(s). We address these gaps here. For each setting, we propose novel estimands, identify them from the observed data, and propose a robust and efficient estimator based on solving the efficient influence function (EIF) estimating equation. 

This paper is organized as follows. We give notation in Section 2 and describe the two structural causal models we consider---a single-instrument scenario and a two-related-instrument scenario. In Section 3 we define complier direct and indirect effects in each scenario and give their identification results. In Section 4, we propose an efficient and robust estimator for each estimand using flexible, data-adaptive regression methods. Section 5 details a limited simulation study evaluating the finite sample performance of these estimators under several scenarios. In Section 6, we apply each estimator to the MTO motivating example to estimate the complier direct and indirect effects in both the single-instrument and double-instrument scenarios. These result in different definitions of ``compliers'' and thus, different estimands. Section 7 concludes.

\section{Notation and structural causal models}
We consider two structural causal models (SCMs). 

\subsection{Single instrument for $Z$}
\label{sec:singleinstrument}
In the first SCM we consider, let $O = (W, A, Z, M, Y)$ represent
the observed data, and let $O_1, \ldots, O_n$ denote a sample of $n$
i.i.d.~observations of $O$. 
Let $W$ denote a vector of observed baseline covariates, $W=f(U_W)$, where $U_W$ is unobserved exogenous error on $W$ and where function $f$ is assumed deterministic but unknown \citep{Pearl2009}. Let $A$ denote a binary instrumental variable \citep{Angristetal&Imbens&Rubin96} of a single, binary exposure variable, $Z$, where $A=f(W,U_A)$ and $Z=f(W, A, U_Z)$ and $U_A$ and $U_Z$ are unobserved exogenous errors. Let $M$ denote a set of mediating variables, $M=f(W, Z, U_M)$, that may be multiple, multi-valued, and/or continuous. Finally, let $Y$ denote a continuous or binary outcome, $Y=f(W,Z,M,U_Y).$ Note that there is no direct effect from $A$ to $M$ or from $A$ to $Y$ in this SCM---$A$ only influences subsequent variables through $Z$. 



\subsection{Related instruments for $Z$ and $M$}
In the second SCM we consider, let $O = (W, A, Z, L, M, Y)$ represent
the observed data, and let $O_1, \ldots, O_n$ denote a sample of $n$
i.i.d.~observations of $O$. 
 Define $W$, $A$, and $Z$ as in Section \ref{sec:singleinstrument}. Let $L$ denote a binary instrumental variable \citep{Angristetal&Imbens&Rubin96} of a single, binary mediator variable, $M$, where $L=f(W,A,Z,U_L)$ and $M=f(W, Z,L, U_M)$ and $U_L$ and $U_M$ are unobserved exogenous errors.  Define $Y$ as in Section \ref{sec:singleinstrument}: $Y=f(W,Z,M,U_Y).$ Note that there is no direct effect from $A$ to $M$, nor from $A$ to $Y$, nor from $L$ to $Y$ in this SCM. Thus, we assume $A$ adheres to the exclusion restriction assumption and is an instrument for the total effect of $Z$ on $Y$, and assume that $L$ adheres to the exclusion restriction assumption and is an instrument for the effect of $M$ on $Y$. Note that the first SCM could be considered a special case of this second SCM where instrument $L$ is not used.

\subsection{General notation}
We use $\P$ to
denote the distribution of $O$, and $\P^c$ to denote the distribution
of $(O,U)$. We let $\P$ be an element of the nonparametric statistical
model defined as all continuous densities on $O$ with respect to some
dominating measure $\nu$. Let $\p$ denote the corresponding
probability density function, and $\mathcal{W, A, Z, L, M}$ denote the range
of the respective random variables. We let $\E$ and $\E^c$ denote
corresponding expectation operators, and define
$\P f = \int f(o)\dd \P(o)$ for a given function $f(o)$. We use
$\g(a \mid w)$ to denote the probability mass function of $A=a$
conditional on $W = w$, and $\e(a \mid m, w)$ to denote the
probability mass function of $A=a$ conditional on $(M, W)=(m,w)$. We
use $\mu(m,z,w)$ to denote the outcome regression function
$\E(Y \mid M = m, Z = z, W = w)$ in the single-instrument setting and $\mu(l,z,w)$ to denote the outcome regression function
$\E(Y \mid L = l, Z = z, W = w)$ in the double-instrument setting.  
We use $\q(z \mid a,w)$ and $\rr(z \mid a,m,w)$ to denote the corresponding conditional densities
of $Z$, $\p(l \mid z,w)$ to denote the conditional density of $L$, and $\cc(m \mid l, z, w)$ to denote the conditional density of $M$ under the second SCM.  

We define counterfactual variables in terms of interventions on the
nonparametric structural equation model (NPSEM). For simplicity of notation, we will use the random variable
with its corresponding intervention in the index to denote the
counterfactual. For example, $Y_{Z=1}$ denotes the random variable
$f_Y(W, 1, M, U_Y)$. For any variable $X$, we let $G_{X=x}$ denote a
random draw from the distribution of the counterfactual
$M_{X=x}$ conditional on $W$. For example, $G_{Z=z}$ is a random draw
from the distribution of $f_M(W, Z=z, U_M)$ conditional on $W$. In this paper we consider interventions that set the mediator equal to such a random draw. This has been referred to as a stochastic intervention on the mediator in the literature \citep{petersen2006estimation,van2008direct,zheng2012targeted, vanderweele2014effect,rudolph2017robust}. In what follows, we drop the random variable from the index
of the counterfactual whenever the variable that is being intervened on is clear from context. For example, we simply use $Y_{a,m}$ to denote
$Y_{A=a,M=m}$, and $M_a$ to denote $M_{A=a}$. 


\section{Definition and identification of (in)direct effects}
We consider a class of mediational effects called population interventional direct and indirect effects, which have been described previously \citep{vanderweele2014effect,vanderweele2017mediation}. We now provide some background on this type of effect. Population interventional total effects can be decomposed into the interventional direct effect and the interventional indirect effect in terms of a contrast between two user-given values $z', z^\star \in Z$ \citep{diaz2019non}: \begin{equation*}
\E^c(Y_{z', G_{z'}} - Y_{z^{\star}, G_{z^{\star}}}) =
      \underbrace{\E^c(Y_{z', G_{z'}} - Y_{z', G_{z^{\star}}})}_{\text{Indirect
          effect (through $M$)}} +
  \underbrace{\E^c(Y_{z', G_{z^{\star}}} - Y_{z^{\star},
      G_{z^{\star}}})}_{\text{Direct effect (not through $M$)}}
\label{eq:decomp}.
\end{equation*} Interventional direct and indirect effects target a population-level path from $Z$ to $Y$ through $M$ for the indirect effect and not through $M$ for the direct effect. If there is no post-instrument exposure, and the assumption $M_0 \perp Y_{1,m} \mid W$ holds, the identification formula (and consequently estimators) of the interventional and natural direct and indirect effects are the same \citep{vanderweele2017mediation}. In what follows we develop a framework for the definition, identification, and estimation of the complier counterparts to these interventional direct and indirect effects.
   
\subsection{Single instrument for $Z$}
\label{sec:defidentsingle}
Under the single-instrument SCM, we have three causal quantities of interest: the complier interventional direct effect (CIDE), the complier interventional indirect effect (CIIE), and the complier interventional total effect (CITE). These are defined as:
\begin{equation}
    \begin{aligned}
        \psi^c_{CIDE}&=\E^c(Y_{Z=1,M=G_{Z=0}} - Y_{Z=0,M=G_{Z=0}} \mid C_Z=1)\\
        \psi^c_{CIIE}&=\E^c(Y_{Z=1,M=G_{Z=1}} - Y_{Z=1,M=G_{Z=0}} \mid C_Z=1)\\
        \psi^c_{CITE}&=\E^c(Y_{Z=1,M=G_{Z=1}} - Y_{Z=0,M=G_{Z=0}} \mid C_Z=1),\\
    \end{aligned}
\end{equation}
where $C_Z=Z_{A=1} - Z_{A=0}$, and $C_Z=1$ denotes compliers in this SCM, i.e., those who would adhere to the treatment if randomized to receive the treatment and would adhere to ``no'' treatment if randomized to not receive treatment.
In words, the interventional direct effect among compliers is the average difference in expected outcomes if $Z$ had been set to 1 versus 0 and stochastically drawing $M$ from the counterfactual joint distribution of mediator values, conditional on $W$, in a hypothetical world in which $Z=0$, among compliers. 
The interventional indirect effect among compliers would analogously be the average difference in expected outcomes setting $Z=1$ and stochastically drawing $M$ from the counterfactual joint distribution of mediator values, conditional on $W$, in a hypothetical world in which $Z=1$ versus $Z=0$, among those who would comply with the intervention. 

Each of the above causal quantities is identified by a statistical parameter that maps a probability distribution $\P$ in the statistical model to a real number. For example, under the following theorem, $\psi^c_{CIDE}$ is identified by the statistical parameter $\psi_{CIDE} = \psi_{IDE} / \psi_{FS}$, where $\psi_{IDE}$ is the statistical parameter for the interventional direct effect of $A$ on $Y$ not through $M$, and where $\psi_{FS}$ is the statistical parameter for the first-stage (FS) effect of instrument $A$ on $Z$. 
 \begin{theorem}\label{theo:single}
 Under the following assumptions, we can identify the complier interventional direct and indirect effects as follows.  
 \begin{assumptioniden}[Exclusion restriction] Assume the exclusion restriction of the NPSEM in which there is no direct effect of $A$ on $M$ and no direct effect of $A$ on $Y$.\label{ass:exclr}
\end{assumptioniden}

\begin{assumptioniden}[Monotonicity] $\P(Z_{A=1}-Z_{A=0}\geq 0) = 1$.\label{ass:mono} Assume that the instrument cannot discourage treatment uptake.
\end{assumptioniden}

\begin{assumptioniden}[Sequential randomization] Assume $Y_{a,m}\indep A\mid W$, $Y_{a,m}\indep M\mid
  (A,Z,W)$, $M_a\indep A\mid W$, and $Z_{A=a}\indep A\mid W$.\label{ass:exch} 
\end{assumptioniden}

\begin{assumptioniden}[Positivity of instrument and mediator
  mechanisms] \label{ass:pos} Assume 
  \begin{itemize}
  \item $\p(w)>0$ implies $\p(a'\mid w)>0$ for $a'\in\{0,1\}$
  \item $\p(m\mid a^\star, w)>0$ and $\p(z\mid a', w)>0$ implies
    $\p(m\mid z,w)>0$  for $(a', a^\star)\in\{0,1\}^2$.
  \end{itemize}
\end{assumptioniden}

\begin{assumptioniden}[Non-zero average effect of the instrument on the treatment] Assume $\E^c(Z_{A=1}-Z_{A=0})\ne 0$.\label{ass:nonzeroeffect}
\end{assumptioniden}

 We can identify the complier interventional direct effect, $\psi_{CIDE}^c$, as $\psi_{IDE} / \psi_{FS}$, where $\psi_{IDE} = \theta(1,0) - \theta(0,0)$, and  $\psi_{FS}=\E\{\E(Z \mid w,A=1)\} - \E\{\E(Z \mid w,A=0)\},$ so
 \begin{align*}
     \E^c(Y_{Z=1,G_{Z=0}} - Y_{Z=0,G_{Z=0}} \mid C_Z=1) & \equiv  \frac{\theta(1,0) - \theta(0,0)}{\E\{\E(Z \mid w,A=1)\} - \E\{\E(Z \mid w,A=0)\}}, 
 \end{align*}
 where 
 $\theta(a', a^\star) = \int\E(Y \mid w,z,m)\dd\P(z \mid a', w)\dd\P(m \mid w,a^\star)\dd\P(w)$. 
 
 We can identify the complier interventional total effect,  $\psi^c_{CITE}$, as $\psi_{ITE} / \psi_{FS},$ where $\psi_{ITE}$ is the
statistical parameter for the interventional total effect of $A$ on $Y$ and $\psi_{ITE}=\theta(1,1) - \theta(0,0),$ so
 \begin{align*}
     \E^c(Y_{Z=1,G_{Z=1}} - Y_{Z=0,G_{Z=0}} \mid C_Z=1) &\equiv \frac{\theta(1,1) - \theta(0,0)}{\E\{\E(Z \mid w,A=1)\} - \E\{\E(Z \mid w,A=0)\}}.
 \end{align*} The complier interventional indirect
  effect may be obtained by subtraction.
 \end{theorem}
 
The proof for the identification result is given in the Supplementary Materials.

Note that we do not assume outcome exchangeability in terms of treatment taken/adherence, $Z$, and therefore allow for unmeasured common causes of $Z$ and $Y$. We also do not require that the instrument is randomized, only that it satisfies exchangeability with respect to the outcome and mediator conditional on $W$.

\subsection{Related instruments for $Z$ and $M$}\label{sec:defidentdouble}
Under the double-instrument SCM, we have two analogous causal quantities of interest: the double complier interventional direct effect (DCIDE) and the double complier interventional indirect effect (DCIIE). These are defined as: \begin{equation}
    \begin{aligned}
        \psi_{DCIDE}^c&=\E(Y_{Z=1,M=G_{Z=0}} - Y_{Z=0,M=G_{Z=0}} \mid C_Z=1, C_M=1)\\
        \psi_{DCIIE}^c&=\E(Y_{Z=1,M=G_{Z=1}} - Y_{Z=1,M=G_{Z=0}} \mid C_Z=1, C_M=1),\\
    \end{aligned}
\end{equation}
where $C_Z=Z_{A=1} - Z_{A=0}=1$ denotes compliers for the instrument for $Z$ and $C_M=M_{L=1}- M_{L=0}=1$ denotes compliers for the instrument for $M$. 

The statistical parameters corresponding to each of these causal quantities can be defined analogously to the definitions in Section \ref{sec:defidentsingle}. For example, under the following theorem, $\psi_{DCIDE}^c$ is identified by the statistical parameter $\psi_{DCIDE}=\psi_{TIIDE} / \psi_{JFS},$ where $\psi_{TIIDE}$ is the statistical parameter for the two-instrument scenario interventional direct effect (TIIDE) of $A$ on $Y$ not through $M$, and where $\psi_{JFS}$ is the statistical parameter for the joint first-stage (JFS) effects of instrument $A$ on $Z$ and instrument $L$ on $M$. 

\begin{theorem}
Under the following assumptions, we can identify the double complier interventional direct and indirect effects as follows. (Compared to Theorem~\ref{theo:single}, we update the exclusion restriction, monotonicity, nonzero average effect of instruments, sequential randomization, and positivity assumptions to relate to both instruments $A$ and $L$.)

\begin{assumptioniden}[Exclusion restriction] Assume the exclusion restrictions of the NPSEM in which there is no direct effect of $A$ on $M$, no direct effect of $A$ on $Y$, and no direct effect of $L$ on $Y$.\label{ass:exclr2}
\end{assumptioniden}

\begin{assumptioniden}[Monotonicity of treatment and mediator instrumentation] Assume $\P(C_Z=C_M\geq 0) = 1$.\label{ass:mono2} 
We note that one could test this assumption in the observed data. 
\end{assumptioniden}

\begin{assumptioniden}[Exchangeability of instruments] Assume $Y_{A=a,L=l}\indep A\mid W$, $Y_{A=a,L=l}\indep L\mid
  (A,Z,W)$, $M_{L=l}\indep L\mid
  (A,Z,W)$, $M_{A=a}\indep A\mid W$, $Z_{A=a}\indep A\mid W$.\label{ass:exch2}
\end{assumptioniden}

\begin{assumptioniden}[Positivity of instrument and mediator
  mechanisms] \label{ass:pos2} Assume 
  \begin{itemize}
  \item $\p(w)>0$ implies $\p(a'\mid w)>0$ for $a'\in\{0,1\}$
  \item $\p(z \mid a^\star,w)>0$ implies $\p(l\mid z, a^\star, w)>0$ for  $a^\star \in\{0,1\}$ and for $l \in \{0,1\}$
  \end{itemize}
\end{assumptioniden}

\begin{assumptioniden}[Non-zero average joint effect of the instruments] 
Assume $\E[(Z_{A=1}-Z_{A=0})(M_{L=1}-M_{L=0})]\ne 0$.
\label{ass:nonzeroeffect2}
\end{assumptioniden}

Define the sequential regression parameters\begin{align*}
\mu(l,z,w) &= \E(Y | L=l, Z=z, W=w)\\
\bar\mu_Z(l,a',w) &= \E[\E(Y\mid L=l, Z, W)\mid A=a', W]\\
    \phi(a,l) &= \E\{\E[Z\E(M\mid L=l, Z, W)\mid A=a, W]\}.
\end{align*}
Note that $\phi(a,l)$ can be rewritten:
\begin{align*} \phi(a,l) &= \E\{\P(M=1 \mid l, Z=1, W)\P(Z=1 \mid a, W) \} 
\end{align*}

Define
\[\gamma(m\mid a, w) = \P(M=m\mid A=a, W),\] and 
\[\vartheta(a',a^\star) =\sum_{l\in\{0,1\}} \int \bar\mu_Z(l,a',w)\gamma(l\mid a^\star,w) \dd\P(w)\]

We can identify the double complier interventional direct effect, $\psi_{DCIDE}^c,$ as $\psi_{TIIDE}/\psi_{JFS},$ where $\psi_{TIIDE}=\vartheta(1,0) - \vartheta(0,0)$, and $\psi_{JFS}=\phi(1,1)-\phi(1,0)-\phi(0,1)+\phi(0,0),$ so
\[\E(Y_{Z=1,
                       M=G_{Z=0}} -
                       Y_{Z=0,
                       M=G_{Z=0}}\mid
     C_Z=1, C_M=1)=\frac{\vartheta(1,0) - \vartheta(0,0)}{\phi(1,1)-\phi(1,0)-\phi(0,1)+\phi(0,0)}.\]
    
We can identify the double complier interventional indirect effect, $\psi_{DCIIE}^c,$ as $\psi_{TIIIE}/\psi_{JFS},$ where $\psi_{TIIIE}=\vartheta(1,1) - \vartheta(1,0)$, so
\[\E(Y_{Z=1,
                       M=G_{Z=1}} -
                       Y_{Z=1,
                       M=G_{Z=0}}\mid
     C_Z=1, C_M=1)=\frac{\vartheta(1,1) - \vartheta(1,0)}{\phi(1,1)-\phi(1,0)-\phi(0,1)+\phi(0,0)}.\]
\end{theorem}

The proof for the identification result of  $\psi_{DCIIE}^c$ is given in the Supplementary Materials.\\
Note that the above identification allows for the decomposition of the double complier interventional total effect (DCITE) into the double complier interventional direct effect (DCIDE) and the double complier interventional indirect effect (DCIIE). If we do not need the 
decomposition, and only the DCIDE is of interest, we can weaken the identification assumptions as in the following theorem. 

\begin{theorem}
Under assumptions \ref{ass:mono}, \ref{ass:nonzeroeffect}, \ref{ass:exclr2}, \ref{ass:pos2} and the following exchangability assumption:
\begin{assumptioniden}[Exchangeability of instruments] Assume $Y_{A=a,L=l}\indep A\mid W$, $Y_{A=a,L=l}\indep L\mid
  (A,Z,W)$, $M_{A=a}\indep A\mid W$, $Z_{A=a}\indep A\mid W$\label{ass:exch3},
\end{assumptioniden}

we can identify the double complier interventional direct effect, $\psi_{DCIDE}^c,$ as $\psi_{TIIDE}/\psi_{FS},$ so
 \[\E(Y_{Z=1,
                       M=G_{Z=0}} -
                       Y_{Z=0,
                       M=G_{Z=0}}\mid
     C_Z=1)=
   \frac{\vartheta(1,0) - \vartheta(0,0)}{ \int [\E(Z \mid w,A=1) - \E(Z \mid w,A=0)]\dd\P(w)}.\]
\end{theorem}
   
\section{Estimation}
We now describe a proposed estimation approach for the complier and double complier interventional direct effects and complier and double complier interventional indirect effects under each of the two SCMs we consider. Under this estimation approach, we estimate the numerator and denominator separately such that we have a ratio of so-called ``one-step'' estimators that each solves its respective EIF estimating equation in one step \citep{Bickel97}.

\subsection{Single instrument for $Z$}
\label{sec:estimsingleIV}
We first consider the one-instrument scenario represented by observed data $O=(W,A,Z,M,Y)$, and the $\psi_{CIDE}$ statistical parameter. We describe how to estimate $\psi_{CIDE}$ by using a one-step estimator of the numerator, $\psi_{IDE}$, and denominator, $\psi_{FS}$, separately. 
An R package to implement this estimator is included: \url{https://github.com/nt-williams/iv_mediation/tree/main/single} 

The EIFs for $\psi_{IDE}$ and $\psi_{FS}$ have been derived in \citet{diaz2019non} and \citet{hahn1998role}, respectively. We provide the formula for each in the Supplementary Materials. Consequently, we can write the EIF for $\psi_{CIDE}$, denoted $D_{CIDE}(O, \P)$, using the delta method:
\begin{equation}
    D_{CIDE}(O,\P) = \frac{D_{IDE}(O, \P)}{\psi_{FS}} - \frac{\psi_{IDE}D_{FS}(O,\P)}{\psi^2_{FS}}.\end{equation}

Let $\eta = (\q, \g, \mu, \rr,\e)$. Let $\hat\eta$ denote an estimator of $\eta$. Let $D_{CIDE}(O,\P)=D_{CIDE}(O,\eta),$ as $\eta$ contains all the relevant features of $\P$. Let 
$D_{DCIDE}(O,\hat{\eta})$ denote an estimator of $D_{DCIDE}(O,\eta)$. For our estimator of $\psi_{CIDE}$, we estimate the numerator and denominator separately, where each is the solution to their respective EIF estimating equation, and then take the ratio of the two. 
    
Note that $\g(a|w)$ and $\q(z|a,w)$ appear in both $D_{FS}(O,\P)$ and $D_{IDE}(O,\P)$, so for compatability, we use the same estimated distribution. 
We also use a cross-fitted version of this estimator. Cross-fitting is a data-splitting technique recently proposed for causal inference with the goal of waeakening some of the technical assumptions required for asymptotic normality of the resulting estimators \citep{klaassen1987consistent,zheng2011cross, chernozhukov2016double}. We perform crossfitting for estimation of all the components of $\eta$ as follows. Let ${\cal V}_1, \ldots, {\cal V}_J$
denote a random partition of data with indices $i \in \{1, \ldots, n\}$ into $J$
prediction sets of approximately the same size such that 
$\bigcup_{j=1}^J {\cal V}_j = \{1, \ldots, n\}$. For each $j$,
the training sample is given by
${\cal T}_j = \{1, \ldots, n\} \setminus {\cal V}_j$. 
$\hat \eta_{j}$ denotes the estimator of $\eta$, obtained by training
the corresponding prediction algorithm using only data in the sample
${\cal T}_j$, and $j(i)$ denotes the index of the
validation set which contains observation $i$. We then use these fits, $\hat\eta_{j(i)}(O_i)$ in computing each efficient influence function, i.e., we compute $D(O_i, \hat\eta_{j(i)})$. 

This estimator of $\psi_{CIDE}$ can be calculated in the following steps:
\begin{enumerate}
    \item  Let the components of $\eta$ be defined as above. Each can be estimated by fitting a regression of the dependent variable on the independent variables and generating predicted probabilities (if the dependent variable is binary) or predicted values (if the dependent variable is numerical), setting the values of independent variables where indicated. For example, $\hat\g(a' \mid w)$ can be estimated by fitting a logistic regression model of $A$ on $W$ and generating predicted probabilities that $A=a'$ for all observed $w$.  One could also use machine learning in model fitting, which is what we do in the simulations and data analysis.
    \item The function $\uu(z,w)$ can be estimated by regressing the quantity $\mu(Z,M,W) \times \h(Z,M,W)$ on $Z,W$ and getting predicted values, setting $(z,w)=(Z,W)$. The function $\vv(w)$ can be estimated by marginalizing out $z$ from $\mu(z,m,w)$ using $\q(z\mid a^{\prime},w)$ as predicted probabilities for each $z$, and then regressing the resulting quantity on $A,W$ and predicting values setting $(a^*,w)=(A,W)$.
    \item The estimate of the numerator, $\psi_{IDE}$, is given by solving $\frac{1}{n}\sum_{i=1}^n D_{IDE}(O_i, \hat{\eta}_{j(i)}) = 0,$ where the components of $\hat{\eta}_{j(i)}$ are estimated as described in Step 1. 
    \item The estimate of the denominator, $\psi_{FS}$, is given by solving $\frac{1}{n}\sum_{i=1}^n D_{FS}(O_i, \hat{\g}_{j(i)},\hat{\q}_{j(i)}) = 0,$ where $\hat{\g}_{j(i)}$ and $\hat{\q}_{j(i)}$ are estimated as described in Step 1.
    \item The ratio of these two estimates gives the one-step estimate of $\psi_{CIDE}$, which we denote with $\hat\psi_{CIDE}$.
    \item The variance of $\hat\psi_{CIDE}$ can be estimated as the sample variance of $D_{CIDE}(O_i;\hat{\eta}_{j(i)})$.
\end{enumerate}

The complier total interventional effect can be estimated analogously, and the complier interventional indirect effect can then be estimated by subtracting the indirect effect from the total effect. 

\begin{lemma}[Double robustness of single-instrument estimator]\label{lemma:singleIVrobust}
  Let $\hat\eta=(\hat\q, \hat\g, \hat\mu, \hat\rr, \hat\e)$ converge to some $\eta_1$ in $L_2(\P)$ norm, and let $\eta_1$ be such that one
  of the following conditions hold:
  \begin{enumerate}[label=(\roman*)]
  \item $(\g_1, \q_1,\e_1,\rr_1)=(\g, \q,\e,\rr)$, or
  \item $(\g_1, \q_1,\mu_1)=(\g, \q,\mu)$.
  \end{enumerate}
  Then $\hat\psi_{CIDE}$ is consistent.
\end{lemma}
Lemma~\ref{lemma:singleIVrobust} implies that it is possible to construct consistent
estimators for $\psi_{CIDE}$ under consistent estimation of the nuisance parameters
in $\eta$ in the configurations described therein. The robustness conditions result from the intersection of robustness of the numerator and robustness of the denominator, which follow from results given in \citet{diaz2019non} and \citet{hahn1998role}.

Asymptotic normality and efficiency of $\hat\psi_{IDE}$ and $\hat\psi_{FS}$ has been previously established \citep{zheng2011cross,diaz2019non}. The conditions for asymptotic normality include convergence of all the components $\hat\eta$ in $L_2(\P)$ norm to their true counterparts at $n^{-1/4}$-rate or faster. A straightforward application of the Delta method yields asymptotic normality of $\hat\psi_{CIDE}$, with asymptotic variance equal to the variance of $D_{CIDE}(O;\eta)$. This licenses the construction of the Wald-type confidence intervals that we use in our simulation studies and illustrative application.

\subsection{Instruments for $Z$ and $M$}
We next consider the two, possibly related, instrument scenario represented by observed data \\
$O=(W,A,Z, L,M,Y),$ and the double complier interventional direct effect target parameter, $\psi_{DCIDE}$. We describe how to estimate $\psi_{DCIDE}$ by using a one-step estimator of the numerator, $\psi_{TIIDE}$, and denominator, $\psi_{JFS}$, separately. An R package to implement this estimator is included: \url{https://github.com/nt-williams/iv_mediation/tree/main/double}

Recall from Section \ref{sec:defidentdouble} that $\psi_{TIIDE}$ can be written as $\vartheta(1, 0) - \vartheta(0, 0)$. 
Similarly, the EIF for $\psi_{TIIDE}$ can be denoted 
\begin{equation}
    D_{TIIDE}(O;\P) = D_{\vartheta}^{1,0}(O;\P) - D_{\vartheta}^{0,0}(O;\P),
\end{equation} where $D_{\vartheta}^{a', a^\star}(O;\P)$ is the EIF for $\vartheta(a',a^\star)$ (for a given $(a', a^\star) \in \{(1,1), (1,0), (0,0)\}$). 
Since $\vartheta(a', a^\star)$ is a particular case of the parameter discussed in \citet{diaz2019non}, its EIF is also a particular case of the EIF in that reference, namely:
\begin{align}
\begin{split}
\label{eq:eiftiide}
    D_{\vartheta}^{a', a^\star}(O;\P) &= \frac{\one\{A=a'\}\gamma(L\mid a^\star, W)}{\p(L\mid Z, a', W)\g(a'\mid W)}\{Y - \mu(L,Z,W)\}\\
    &+\frac{\one\{A=a'\}}{\g(a'\mid W)}\left\{\bar \mu_L(Z,a^\star,W) - \sum_z \bar \mu_L(z,a^\star,W)\q(z\mid a', W)\right\}\\
    &+\frac{\one\{A=a^\star\}}{\g(a^\star\mid W)}\left\{\bar \mu_Z(M,a',W) - \sum_l \bar \mu_Z(l,a',W)\gamma(l\mid a^\star,W)\right\}\\
    &+\sum_{l\in\{0,1\}}\bar \mu_Z(l,a',W)\gamma(l\mid a^\star,W) - \vartheta(a',a^\star), 
    \end{split}
\end{align}
where
\begin{align*}
\bar\mu_L(Z,a',W) &= \sum_{l\in\{0,1\}}\mu(l,Z,W)\gamma(l\mid a^\star,W),
\end{align*}
and where $\gamma$ can be rewritten as
\begin{align*}
\gamma(m \mid a^\star, W) &= \P(M=m \mid a^\star, W)\\
&= \sum_{l, z} \cc(m | l, z, W) \p(l | z, a^\star, W) \q(z | a^\star, W).
 \end{align*}

The EIF for $\psi_{JFS}$, denoted $D_{JFS}(O;\P)$, is the same as for a longitudinal intervention with two treatment nodes, $A$ and $L$, and time-varying covariate, $Z$, which has been derived by \citet{van2012targeted}. 

We can write the EIF for $\psi_{DCIDE}$, denoted $D_{DCIDE}(O;\P)$, using the delta method.

\begin{equation}
    D_{DCIDE}(O;\P) = \frac{D_{TIIDE}(O;\P)}{\psi_{JFS}} - \frac{\psi_{TIIDE}D_{JFS}(O;\P)}{\psi^2_{JFS}},
\end{equation} where
\begin{equation}
    D_{JFS}(O;\P) = D_{\phi}^{1,1}(O;\P) - D_{\phi}^{1,0}(O;\P) -D_{\phi}^{0,1}(O;\P) +D_{\phi}^{0,0}(O;\P),
\end{equation} and where
\begin{align}
\begin{split}
   D_{\phi}^{a, l}(O;\P) &=  \frac{I(A=a, L=l)}{\g(a \mid W)\p(l \mid Z, a, W)}(MZ - Z\cc(M=1 \mid l, Z, W) \\
   &+ \frac{I(A=a)}{\g(a \mid W)}(Z\cc(M=1 \mid l, Z, W) - \cc(M=1 \mid l, Z=1, W)\q(Z=1 \mid a,W)) \\
   &+ \cc(M=1 \mid l, Z, W)\q(Z=1 \mid a,W)- \phi(a,l)
   \end{split}
\end{align} 

Let $\kappa=(\g, \q, \p, \cc, \mu)$ represent the set of nuisance parameters used in $D_{DCIDE}(O,\P).$ Let $\hat{\kappa}$ denote an estimator of $\kappa$. Let $D_{DCIDE}(O,\P)=D_{DCIDE}(O,\kappa),$ as $\kappa$ contains all relevant features of $\P$. Let $D_{DCIDE}(O,\hat\kappa)$ denote an estimator of $D_{DCIDE}(O,\kappa)$. For our estimator of $\psi_{DCIDE}$, we estimate the numerator and denominator separately, where each is the solution to their respective EIF estimating equation, and then take the ratio of the two.
 
Note that $\g(a|w)$, $\q(z|a,w)$, $\p(l|z,a,w)$, and $\cc(m|l,z,w)$ appear in both $D_{JFS}(O,\P)$ and $D_{TIIDE}(O,\P)$, so for compatability, we use the same estimated distribution. 
We also use a cross-fitted version of this estimator \citep{klaassen1987consistent,zheng2011cross, chernozhukov2016double}, performing crossfitting in all fits, as described in Section \ref{sec:estimsingleIV}. 

This estimator of $\psi_{DCIDE}$ can be computed in the following steps:
\begin{enumerate}
    \item Let the components of $\kappa$ be defined as above. Each can be estimated by fitting a regression of the dependent variable on the independent variables where indicated. For example, $\hat{\p}(l' \mid Z, A, W)$ can estimated by fitting a logistic regression model of $L$ on $(Z, A, W)$ and generating predicted probabilities that $L=l'$. One could also use machine learning in model fitting, which is what we do in the simulations and data analysis.
    \item The function $\bar{\mu}_Z(l,a',w)$ can be estimated by first fitting the quantity $\mu(l,z,w)$, and integrating out $z$, which adds dependence on $a'$.
    The random variable $\bar{\mu}_Z(M,a',W)$ can be constructed by evaluating $\bar{\mu}_Z(l,a',w)$ at $(l,w)=(M,W)$.
    The function $\gamma(m \mid a^\star, w)$ can be estimated by first fitting the quantity $\cc(m  \mid l, z, w)$, and integrating out $l$ with respect to $\hat\p(l | z, a^\star, w)$, and then integrating out $z$ with respect to $\hat\q(z | a^\star, W)$, which adds dependence on $a^\star$. We then compute $\gamma(L \mid a^\star, W)$ by evaluating this function at $(m, w)=(L, W)$.
    The function $\bar{\mu}_L(z,a^\star,w)$ can be estimated by first fitting the quantity $\mu(l,z,w)$, and integrating out $L$, using $\gamma(l \mid a^\star,w)$ which adds dependence on $a^\star$.
    \item The estimate of $\psi_{TIIDE}$ is obtained by solving $\frac{1}{n}\sum_{i=1}^n D_{TIIDE}(O_i, \hat\kappa_{j(i)}) = 0.$ 
    \item The estimate of $\psi_{JFS}$ is given by solving $\frac{1}{n}\sum_{i=1}^n D_{JFS}(O_i, \hat\g_{j(i)}, \hat\q_{j(i)}, \hat\p_{j(i)}, \hat\cc_{j(i)}) = 0,$ where $\hat\g_{j(i)}, \hat\q_{j(i)}, \hat\p_{j(i)}, \hat\cc_{j(i)}$ are estimated as described in Step 1 and using cross-fitting.
    \item The ratio of these two estimates gives the one-step estimate of $\psi_{DCIDE}$, which we denote with $\hat\psi_{DCIDE}$.
    \item The variance can be estimated as the sample variance of $D_{DCIDE}(O_i;\hat\kappa_{j(i)})$.
\end{enumerate}

The double complier interventional indirect effect and double complier interventional total effect can then be estimated analogously. 

\begin{lemma}[Double robustness of double-instrument estimator]\label{lemma:doubleIVrobust}
  Let $\hat\kappa=(\hat\q, \hat\g, \hat\mu, \hat\p, \hat\cc)$ converge to some $\kappa_1$ in $L_2(\P)$ norm, and let $\kappa_1$ be such that one
  of the following conditions hold:
  \begin{enumerate}[label=(\roman*)]
   \item $(\q_1, \mu_1, \cc_1)=(\q,\mu,\cc)$, or
  \item $(\g_1, \q_1,\mu_1, \p_1)=(\g, \q,\mu,\p)$, or
  \item $(\g_1, \q_1, \p_1, \cc_1)=(\g, \q,\p,\cc)$, or
  \item $(\g_1, \mu_1, \p_1, \cc_1)=(\g, \mu,\p,\cc)$.
  \end{enumerate}
  Then $\hat\psi_{DCIDE}$ is consistent.
\end{lemma}
Lemma~\ref{lemma:doubleIVrobust} implies that it is possible to construct consistent
estimators for $\psi_{DCIDE}$ under consistent estimation of the nuisance parameters
in $\kappa$ in the configurations described therein. The robustness conditions result from the intersection of robustness of the numerator and robustness of the denominator. As in the single instrument case, $\hat\psi_{DCIDE}$ is asymptotically normal under conditions which include the convergence of each component of $\hat\kappa_{j(i)}$ to its true value at $n^{1/4}$-rate in $L(\P)$ norm.

\section{Simulation}
We conducted a limited simulation study to illustrate the properties 
of the one-step estimators in the single-instrument setting and double-instrument setting. However, we acknowledge that these simulations are not informative of general estimator performance. We consider estimator performance in terms of absolute bias, absolute bias scaled by $\sqrt{n}$, mean squared error relative to the efficiency bound scaled by ${n}$ and 95\% confidence interval (CI) coverage. We conducted 1000 simulations for sample sizes $n \in \{500, 1000, 2000, 5000\}$. We also consider several specifications of $\eta$ and $\kappa$. 

\subsection{Single-instrument setting}
In the single-instrument setting, we consider the following data-generating mechanism (DGM). All variables are Bernoulli distributed with probabilities given by

{\footnotesize
  \begin{align*}
    P(W_1 = 1) &= 0.6 \\
    P(W_2 = 1) &= 0.3 \\
    P(W_3 = 1 \mid W_1, W_2) &= 0.2 + (W_1 + W_2)/3 \\
    P(A = 1) &= 0.5 \\
    P(Z = 1 \mid A, W) &= \expit(-\log(1.1)*(W_1 + W_2 + W_3)/3 + 3A) \\
    P(M = 1 \mid Z, W) &= \expit(-\log(3)*(W_1 + W_2) + 2Z) \\
    P(Y = 1 \mid M, Z, W) &= \expit(1 / (-W_1 - W_2 - W_3 + Z + M -1) )         
  \end{align*}
 }

$A$ is randomly assigned and adheres to the exclusion restriction \citep{Angristetal&Imbens&Rubin96}, aligned with its role as an instrumental variable. We consider three model specifications: one with all the components of $\eta$ correctly specified, one with $\mu$ inconsistently estimated, and with $(\rr,\e)$ inconsistently estimated. For correct specifications of $(\q, \rr, \e)$, we fit nuisance parameters using $\ell_1$-regularized logistic regression including all interactions \citep{benkeser2016highly,van2017generally}; $\g$ was correctly specified using an intercept-only model and $\mu$ with a main-effects generalized linear model. Inconsistent nuisance parameters are fit using intercept-only models.

\begin{figure}[H]
\caption{Performance of one-step estimator in a single-instrument setting across different model specifications, in terms of (a) scaled bias, (b) 95\% CI coverage, and (c) efficiency compared to the efficiency bound.}
\label{fig:single-perf}
\hfill
\subfloat[$\sqrt{n}$-bias \label{fig:single-bias}]{
    \includegraphics[width=\textwidth]{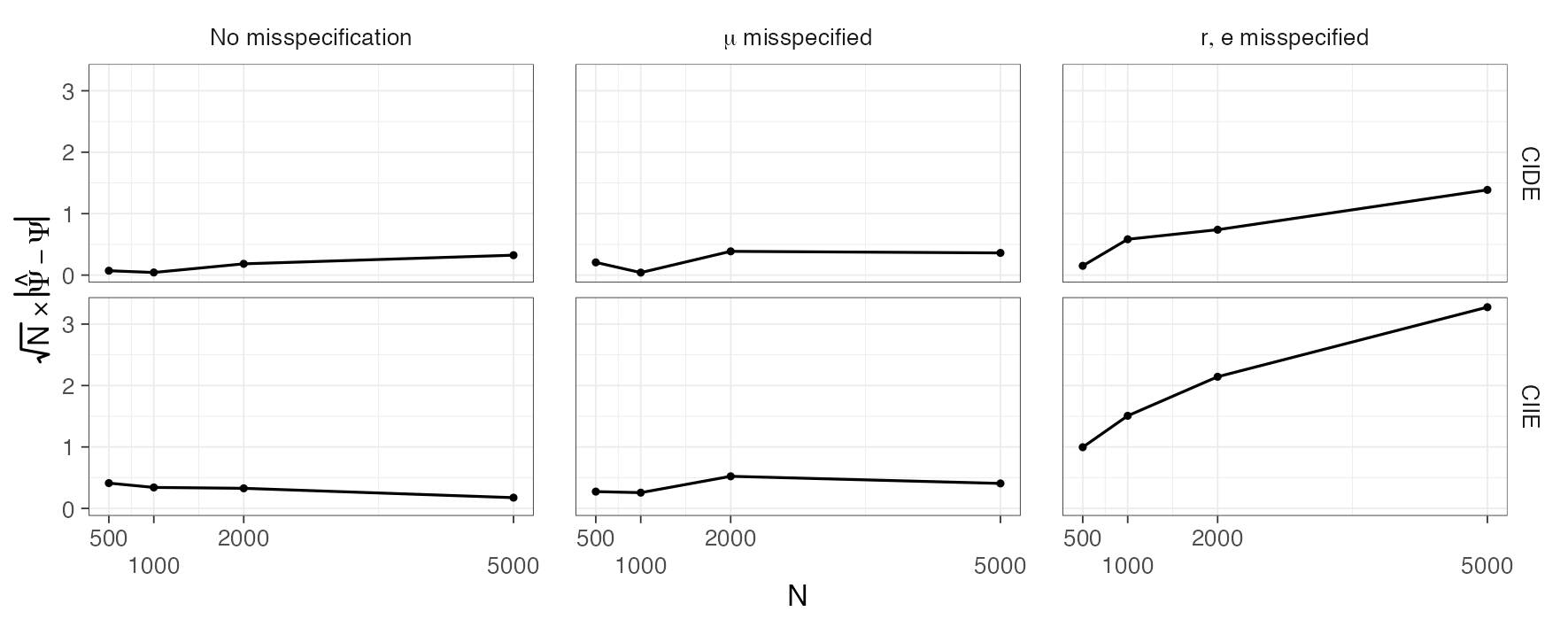}
}
\hfill
\subfloat[95\% CI Coverage \label{fig:single-ci}]{
    \includegraphics[width=\textwidth]{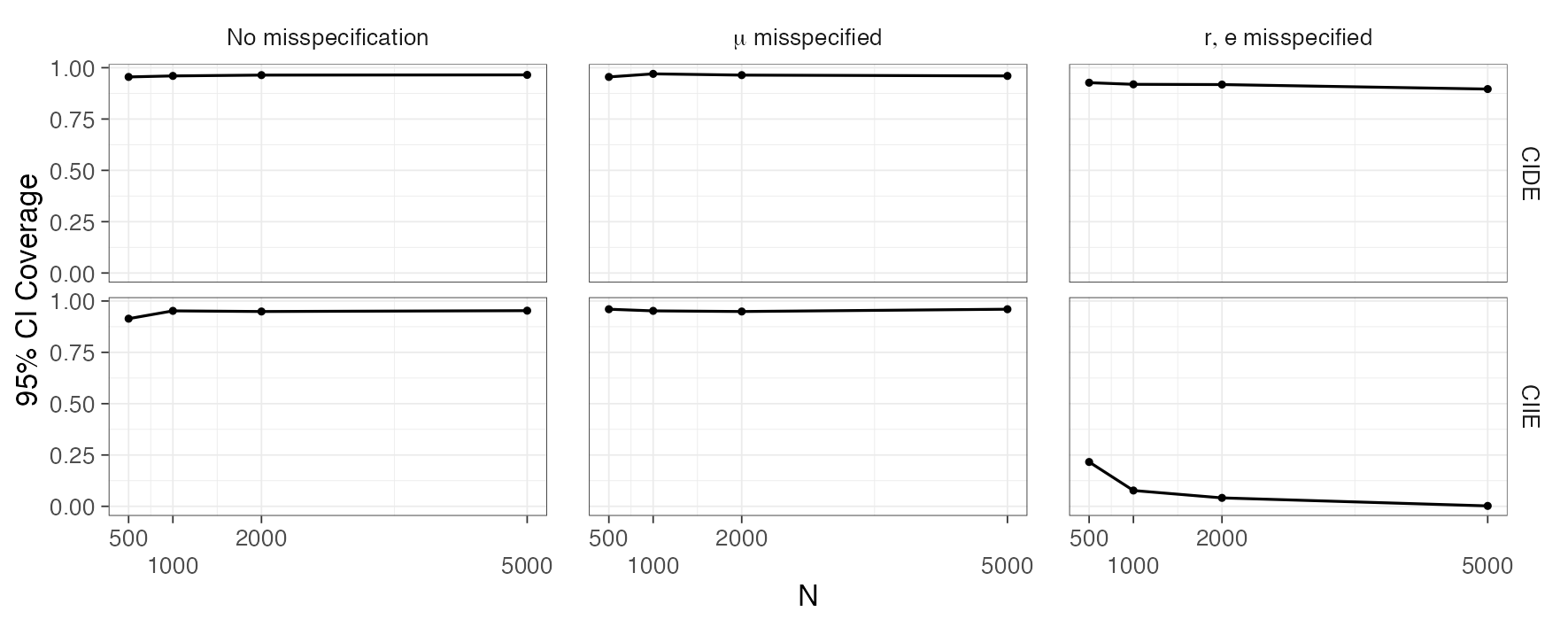}
}
\hfill
\subfloat[Relative efficiency \label{fig:single-rel}]{
    \includegraphics[width=\textwidth]{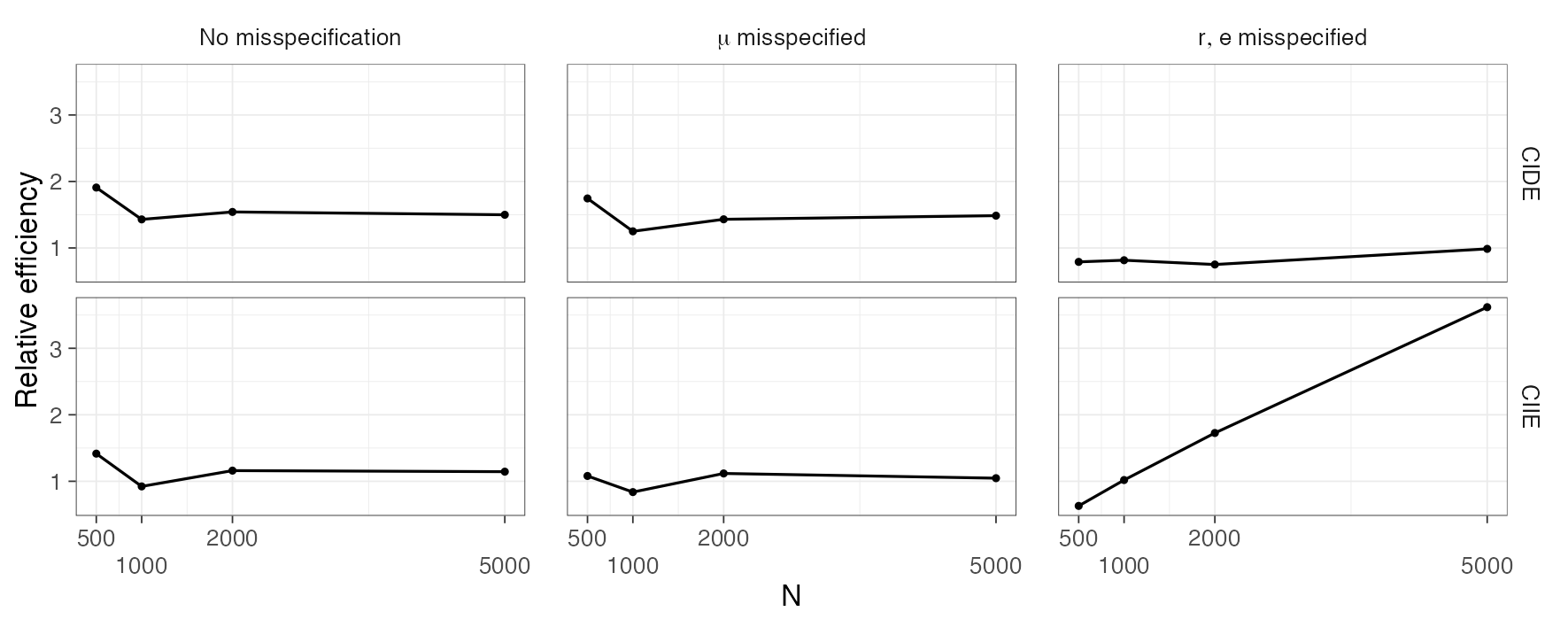}
}
\end{figure}

Figure \ref{fig:single-perf} shows the results of our simulation study for the one-step estimator in a single-instrument setting. When all nuisance parameters are correctly specified, the scaled mean squared error converges to the efficiency bound (Figure \ref{fig:single-rel}) and confidence interval coverage is close to 95\% (Figure \ref{fig:single-ci}). As predicted by  Lemma \ref{lemma:singleIVrobust}, we observe consistent estimates for all specifications in Figure \ref{fig:single-perf}. Contrary to what is expected, close to nominal confidence interval coverage was observed in some of the misspecified simulation scenarios. This should not be expected in general other than when all nuisance parameters are consistently estimated.

\subsection{Double-instrument setting}
In the double-instrument setting, we consider the following DGM. All variables are Bernoulli distributed with probabilities given by

{\footnotesize
  \begin{align*}
    P(W_1 = 1) &= 0.6 \\
    P(W_2 = 1) &= 0.3 \\
    P(W_3 = 1 \mid W_1, W_2) &= 0.2 + (W_1 + W_2)/3 \\
    P(A = 1) &= 0.5 \\
    P(Z = 1 \mid A, W) &= \expit(-\log(1.1)*(W_1 + W_2 + W_3)/3 + 3A) \\
    P(L = 1 \mid Z, W) &= \expit(-\log(2)*(W_1 + W_2 + W_3)/3 + 3Z - 1) \\
    P(M = 1 \mid L, Z, W) &= \expit(-\log(3)*(W_1 + W_2) + 3L + Z - 1) \\
    P(Y = 1 \mid M, Z, W) &= \expit(0.3 - \log(5)*(W_1 + W_2 + W_3) + Z + M)
  \end{align*}
 }

Similar to the single-instrument setting, $A$ is randomly assigned and adheres to the exclusion restriction \citep{Angristetal&Imbens&Rubin96} definition of an instrumental variable. We consider five model specifications: one with all the components of $\kappa$ correctly specified, and where each of $(\mu, \p, \cc, \q)$ are inconsistently estimated. For correct specifications, we fit nuisance parameters using $\ell_1$-regularized logistic regression including all interactions \citep{benkeser2016highly,van2017generally}, except for $\g$ which used an intercept-only model. Incorrect nuisance parameter specifications are fit using an intercept-only model.

\begin{figure}[H]
\caption{Performance of one-step estimator in a double-instrument setting across different model specifications, in terms of (a) scaled bias, (b) 95\% CI coverage, and (c) efficiency compared to the efficiency bound.}
\label{fig:double-perf}
\hfill
\subfloat[$\sqrt{n}$-bias \label{fig:double-bias}]{
    \includegraphics[width=\textwidth]{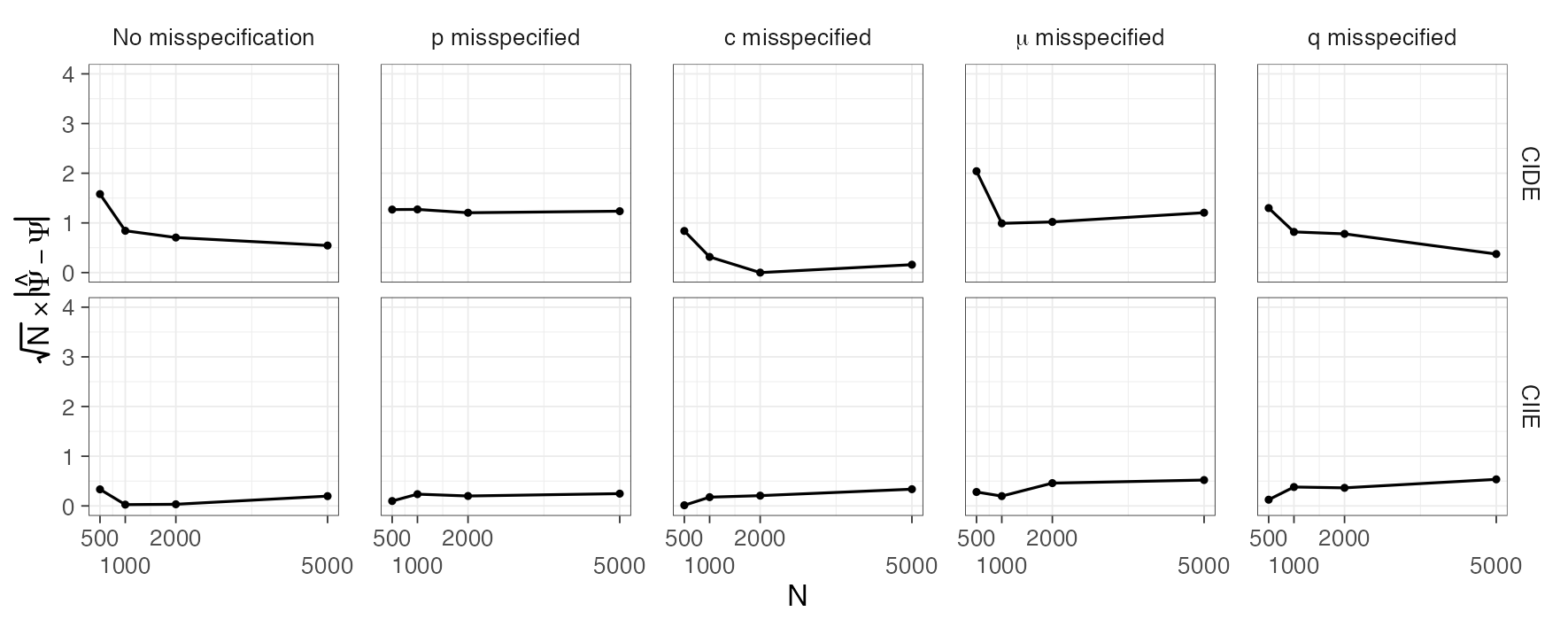}
}
\hfill
\subfloat[95\% CI Coverage \label{fig:double-ci}]{
    \includegraphics[width=\textwidth]{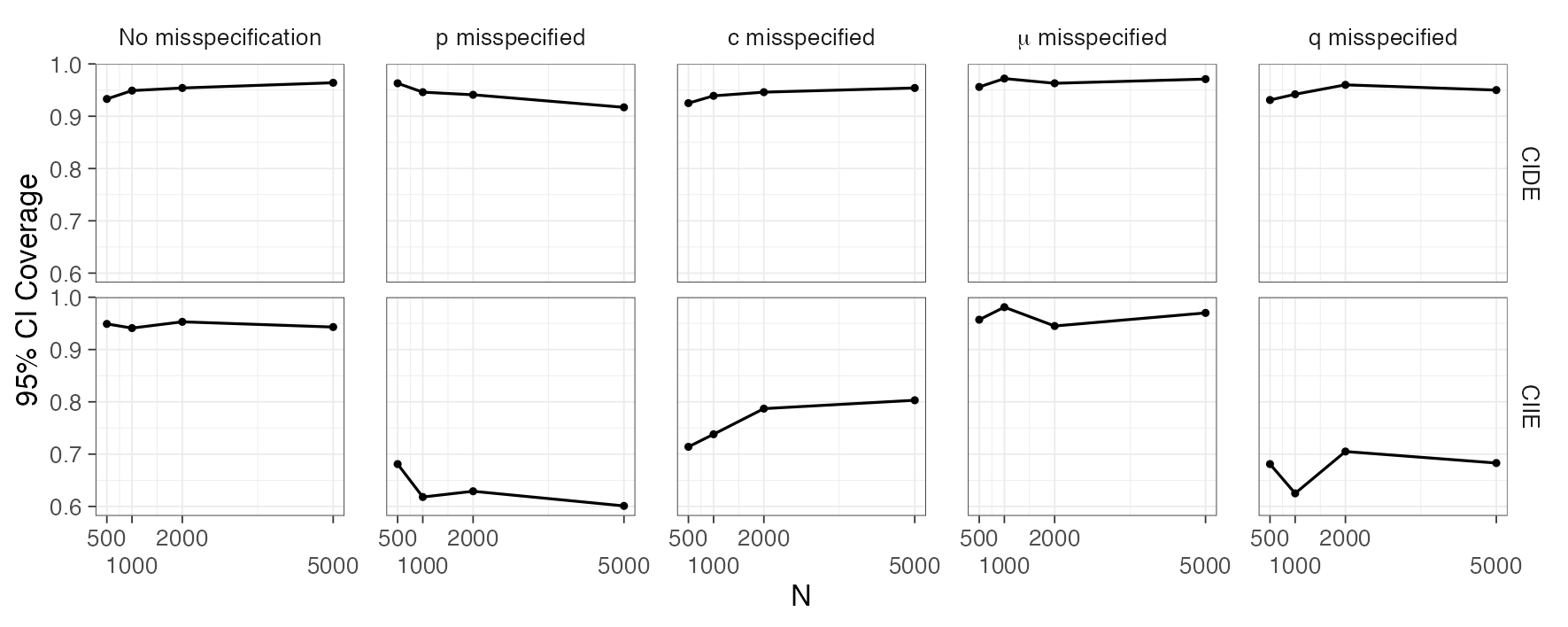}
}
\hfill
\subfloat[Relative efficiency \label{fig:double-rel}]{
    \includegraphics[width=\textwidth]{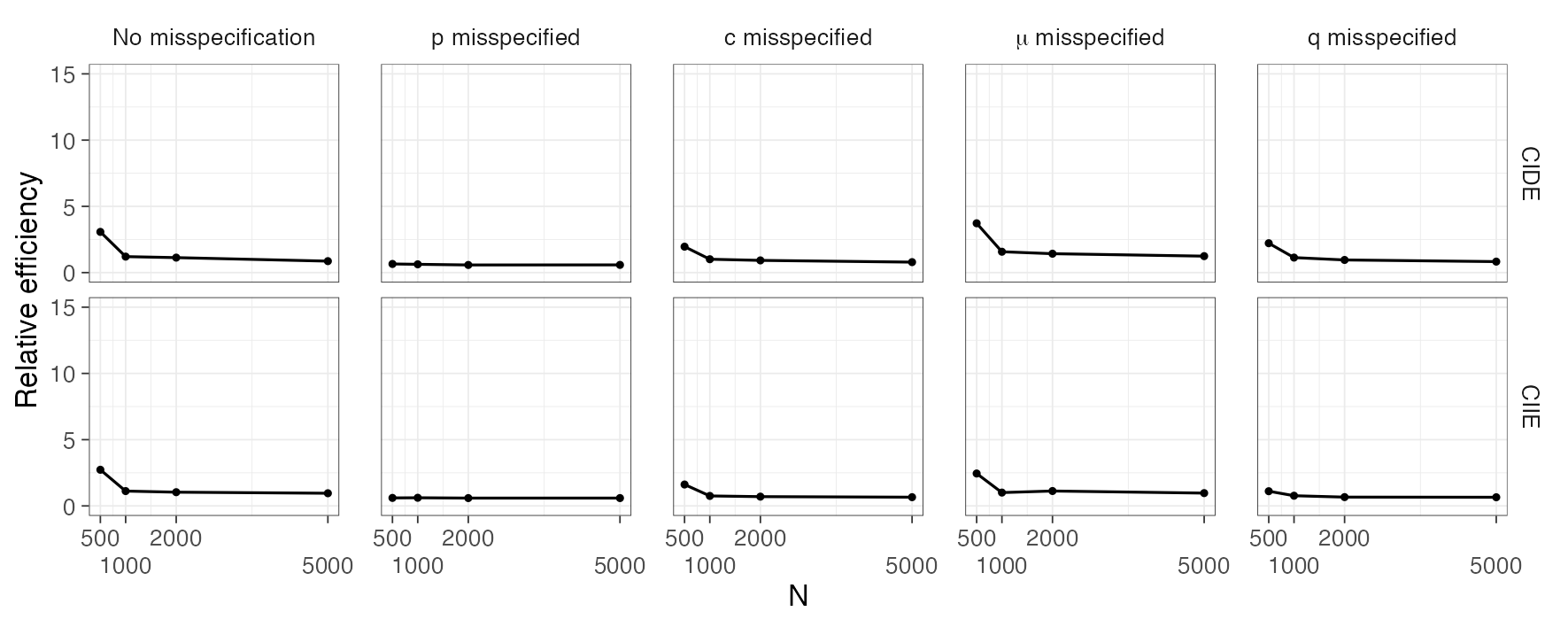}
}
\end{figure}

Figure \ref{fig:double-perf} shows the results of our simulation study for the one-step estimator in a double-instrument setting. As expected, when all nuisance parameters are correctly specified, the scaled mean squared error converges to the efficiency bound (Figure \ref{fig:double-rel}) and confidence interval coverage is close to 95\% (Figure \ref{fig:double-ci}). Given the results of Lemma \ref{lemma:doubleIVrobust}, we observe consistent estimates for all specifications in Figure \ref{fig:double-perf}. Similarly to the single-instrument simulation, close to 95\% confidence interval coverage was observed in many of the simulation scenarios. Once again, this should not be expected other than when all nuisance parameters are consistently estimated.

\section{Empirical Illustration}

We now apply our proposed one-step estimators to two scenarios from the Moving to Opportunity (MTO) study---one that assumes the presence of a single instrument and one that assumes the presence of two instruments. 

\subsection{Background and Sample} MTO was a longitudinal, randomized trial where families living in public housing in five US cities were randomized to receive a Section 8 housing voucher that they could then use to move out of public housing and rent on the private market \citep{kling2007experimental}. Aligned with prior analyses estimating direct and indirect effects in MTO \citep{rudolph2021helped,rudolph2018mediation}, our sample includes adolescents participating in MTO who were 12-17 years old at the final outcome assessment. We exclude the Baltimore site, as Section 8 voucher receipt did not increase a family's likelihood of moving to a low-poverty neighborhood, which differs from other sites and from the intention of the intervention. We conducted analyses among boys only, as previous work documented qualitatively and quantitatively different intervention effects between girls and boys \citep{orr2003moving,clampet2011moving}, with boys experiencing unintended harmful total effects of the MTO intervention on mental health outcomes \citep{kling2007experimental,sanbonmatsu2011moving,kessler2014associations}. Identifying contributing mediation mechanisms can help shed light on why and how such harmful total effects exist \citep{rudolph2021helped,smith2021mediating}. 

\subsection{Data}
In both the single and double instrument scenarios, the instrument, $A$, for the exposure of using the voucher to move out of public housing, is a binary variable defined as randomization to receive a Section 8 housing voucher that one can then use to rent on the private market. The exposure, $Z$, is a binary variable defined as adherence to the intervention---using the housing voucher, if one received it, to move out of public housing. The outcome, $Y$, is another binary variable defined as presence of a psychiatric mood disorder (major depressive disorder or an anxiety disorder, defined based on the Diagnostic Statistical Manual, version IV (DSM-IV)) in the past-year, as measured by the CIDI-SF instrument \citep{kessler1998world}. We include numerous baseline covariates related to the individual, the individual's family, and characteristics of the individual's neighborhood at baseline, $W$. A full list is available in the Supplementary Materials. 

In the single instrument scenario, we have observed data $O=(W, A, Z, M, Y)$. Variables $W, A, Z, Y$ are defined as above. The mediator, $M$, is defined as neighborhood poverty (proportion of neighborhood residents living under the poverty line) after randomization and prior to outcome ascertainment, duration weighted (bounded $[0,1]$). 

In the double instrument scenario, we have observed data $O=(W, A, Z, L, M, Y)$. Variables $W, A, Z, Y$ are defined as above. The instrument for the mediator, $L$, is a binary variable defined as changing school districts after randomization. 
The mediator, $M$, is a binary variable defined as living in low-poverty neighborhoods during follow-up (defined as $<$30\% neighborhood residents living under the poverty line after randomization and prior to outcome ascertainment, duration weighted).

For the purposes of this illustrative example, we use one imputed dataset to address missingness in $W, L, M, $ and $Y$. Variables $A$ and $Z$ have no missingness. 

\subsection{Estimand}
In the example we consider here, the complier interventional direct effect is the direct effect of using the housing voucher to move out of public housing on risk of developing a DSM-IV mood disorder in adolescence, not mediated by subsequent neighborhood poverty, among those who would comply with the intervention. The complier interventional indirect effect is the effect of using the housing voucher to move out of public housing on risk of developing a DSM-IV mood disorder in adolescence, mediated by subsequent neighborhood poverty, among those who would comply with the intervention. We consider both the single-instrument and double-instrument complier interventional direct and indirect effects. In addition, we also consider the standard intent-to-treat interventional direct and indirect effects of being randomized to receive the housing voucher (i.e., non-IV estimands). The intent-to-treat interventional direct effect is the direct effect of being randomized to receive a housing voucher on risk of developing a DSM-IV mood disorder in adolescence, not mediated by subsequent neighborhood poverty. The intent-to-treat interventional indirect effect is the effect of being randomized to receive a housing voucher on risk of developing a DSM-IV mood disorder in adolescence, mediated by subsequent neighborhood poverty. Similar intent-to-treat interventional direct and indirect effects have been considered previously in MTO \citep{rudolph2021helped}.

\subsection{Analysis}
We use the estimators developed herein to estimate each of the above complier and double complier interventional direct and indirect effects. We use previously developed estimators \citep{diaz2019non} to estimate each of the intent-to-treat interventional direct and indirect effects. We use data-adaptive methods
for fitting the nuisance parameters, using the Super Learner
ensembling procedure, which is the convex
combination of user-supplied algorithms that minimizes the
10-fold cross-validated prediction error \citep{van2007super}. These algorithms included
generalized linear models, intercept-only models,
lasso \citep{tibshirani1996regression}, and multivariate adaptive regression splines \citep{friedman1991multivariate}. Standard errors are estimated using the sample variance of the influence curve. Columbia University determined this analysis of deidentified data to be non-human subjects research.

\begin{figure}[H]
\caption{Estimates (and 95\% confidence intervals) of the 1) interventional direct effect of randomized voucher receipt on long-term risk of developing a DSM-IV mood disorder in adolescence, not operating through the mediator neighborhood poverty, and the 2) interventional indirect effect operating through neighborhood poverty; and the 3) complier interventional direct effect of moving with the voucher on long-term risk of developing a DSM-IV mood disorder in adolescence, not operating through the mediator neighborhood poverty, and the 4) complier interventional indirect effect operating through neighborhood poverty. All results were approved for release by the U.S. Census Bureau, authorization number CBDRB-FY22-CES018-004.}
\label{fig:}
\subfloat[Single instrument setting \label{fig:single}]{
  \includegraphics[width=0.50\textwidth]{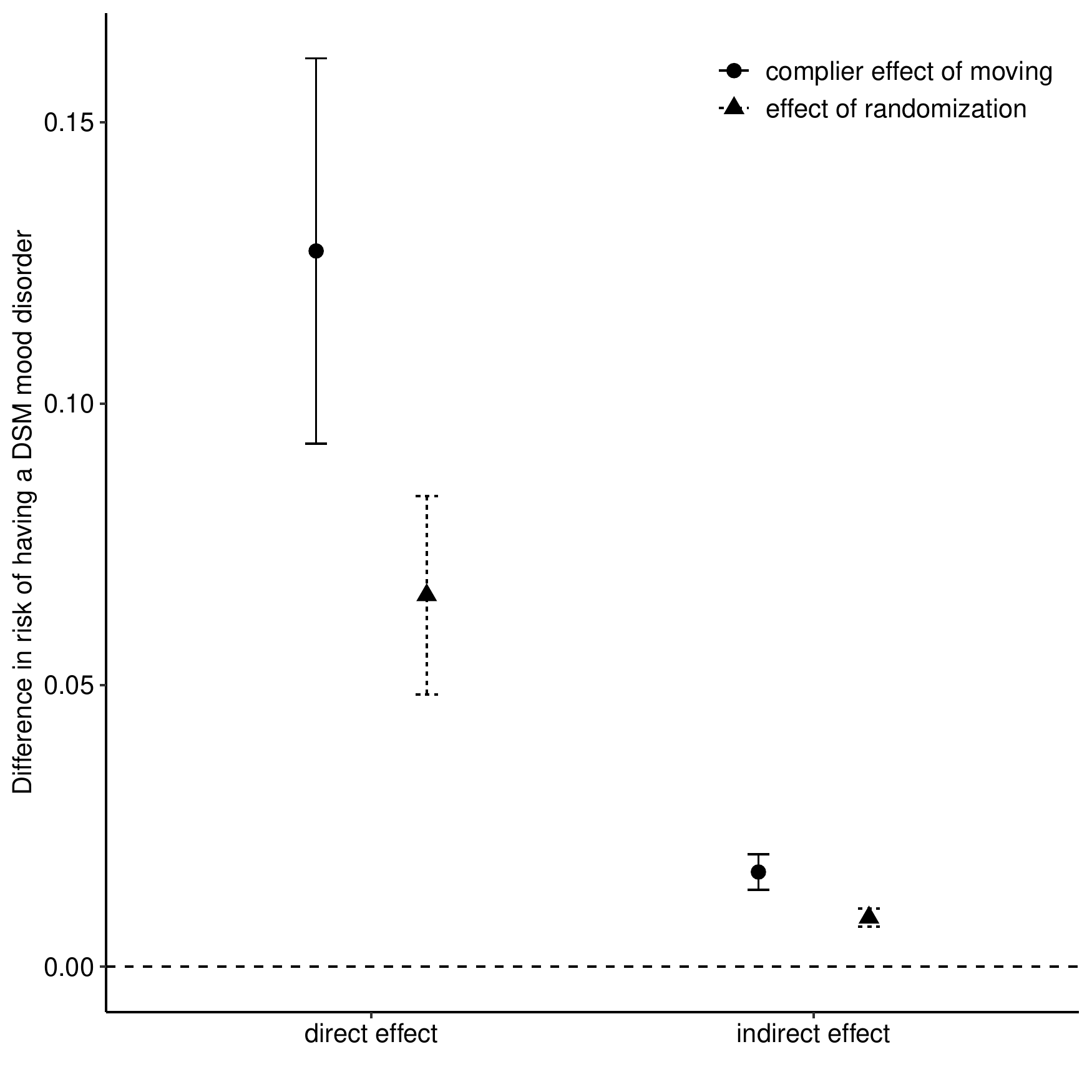}}
     \hfill
\subfloat[Double instrument setting \label{fig:double}]{
\includegraphics[width=0.50\textwidth]{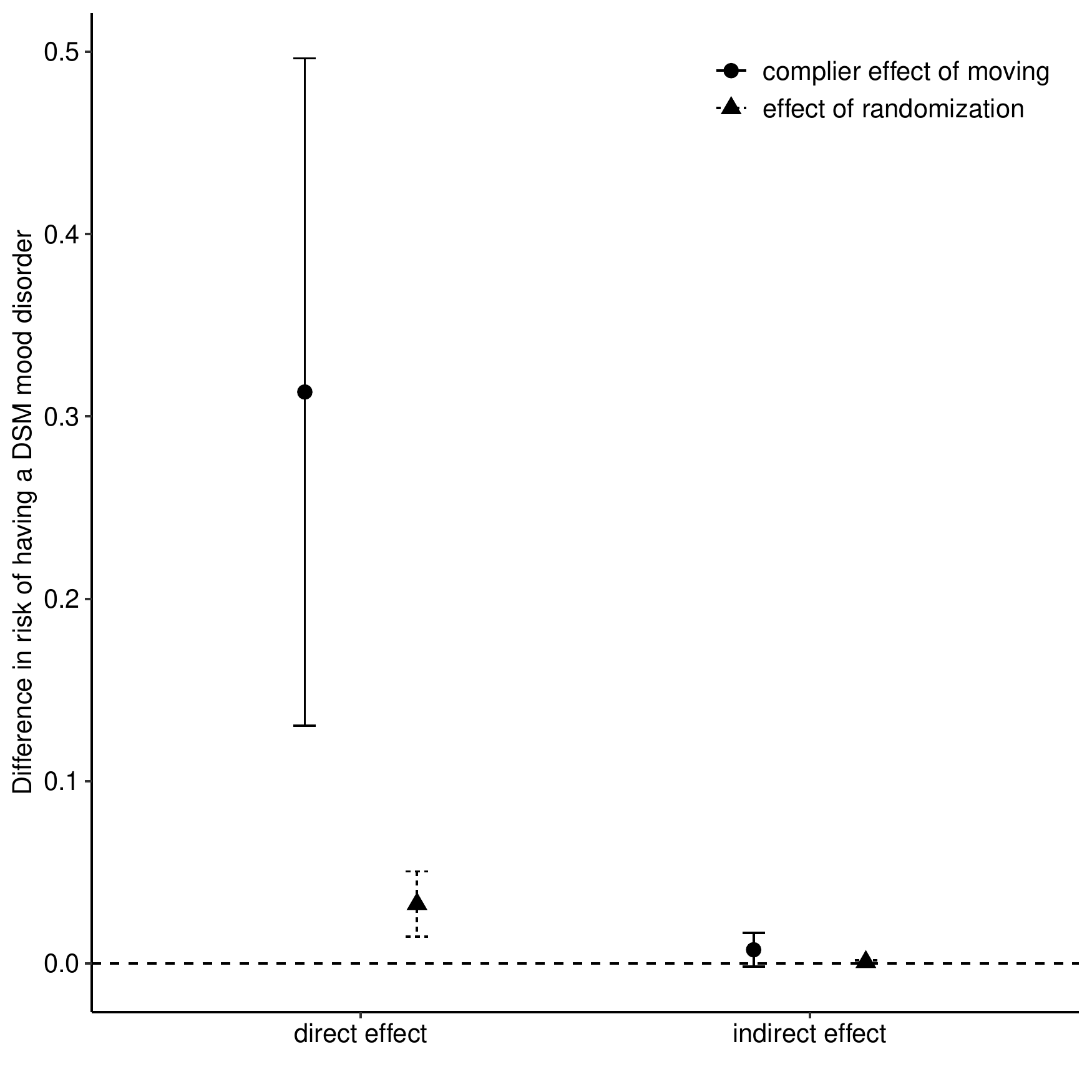}}
\end{figure}

Figure \ref{fig:single} shows the complier interventional direct and indirect effect estimates and 95\% CIs (labeled ``complier effect of moving'') and the intent-to-treat interventional direct and indirect effects and 95\% CIs (labeled ``effect of randomization'') in the single-instrument scenario. Figure \ref{fig:double} shows the double complier interventional direct and indirect effect estimates and 95\% CIs (labeled ``complier effect of moving'') and the intent-to-treat interventional direct and indirect effects and 95\% CIs (labeled ``effect of randomization'')  in the double-instrument scenario. 

In general, we see smaller, more precise effects in the single-instrument setting (Figure \ref{fig:single}). Corroborating previous work \citep{rudolph2021helped}, we see a small, harmful indirect effect of being randomized to receive a voucher on risk of developing a mood disorder among boys that operates through neighborhood poverty as well as a harmful direct effect that does not operate through neighborhood poverty. Also as expected 
the complier direct and indirect effects of moving with the voucher on risk of developing a mood disorder are slightly larger than their intent-to-treat counterparts (complier interventional direct effect (risk difference): 0.127, 95\% CI: 0.093-0.161; complier interventional indirect effect: 0.017, 95\% CI: 0.014-0.020). We would expect complier average effects to be larger than their total intent-to-treat counterparts, because the complier point estimate is calculated by dividing the intent-to-treat estimate by the probability (a number between 0 and 1) of being a complier (i.e., the ``first-stage'' effect of $A$ on $Z$). Estimating effects among the smaller number of compliers could also result in the complier effects having less precision in some cases \citep{jo2002statistical,angristmostlyharmless}. 

The double instrument setting addresses unobserved confounding of the $M-Y$ relationship, but at the expense of precision in terms of both the measure of the mediator (binarized neighborhood poverty) and the resulting estimates (Figure \ref{fig:double}), and, relatedly, at the expense of having a more limited group of double compliers \citep{angristmostlyharmless}. We again see a harmful indirect effect of being randomized to receive a voucher on risk of developing a mood disorder among boys that operates through neighborhood poverty as well as a harmful direct effect that does not operate through neighborhood poverty. These effect estimates are very similar to those in Figure \ref{fig:single}, as expected in the absence of strong unmeasured confounding of the mediator-outcome relationship, as the only difference is whether $M$ is treated as a continuous or binary variable. As in the single-instrument case, the double complier interventional direct and indirect effects of moving with the voucher on risk of developing a mood disorder are also larger and much less precise than the intent-to-treat counterparts (double complier interventional direct effect (risk difference): 0.313, 95\% CI: 0.130-0.496; double complier interventional indirect effect: 0.007, 95\% CI: -0.002-0.017). In this setting, ``double compliers'' are defined as those who 1) were randomized to receive the voucher and moved with it as well as 2) changed school districts, and that change involved a move to a lower poverty neighborhood. The joint additive effect of the two instruments on the exposure and mediator is (0.104, 95\% CI: 0.080-0.128). In other words, being randomized to receive a voucher and changing school districts (conditional on moving, receiving the voucher, and covariates) increases risk of moving with a voucher to a lower-poverty neighborhood by 10.4 percentage points. 

\section{Conclusion}
In this paper, we defined and identified complier interventional direct and indirect effects (i.e., IV mediational effects) in two scenarios: 1) one where there is a single instrument of the exposure of interest, but relies on the assumption of no unobserved confounders of the mediator-outcome relationship, and 2) where there are two instruments, one for the exposure and another for the mediator, that may be related to each other, and thus, no longer relies on an unobserved confounding assumption. We proposed nonparametric, robust, and efficient estimators of each effect, based on solving the EIF estimating equation, described their robustness properties, and evaluated their finite sample performance in a limited simulation study. 

The effects and estimators we propose are widely applicable. Anytime a mediation question is asked in the context of a randomized trial with possible noncompliance, in the context of treatment discontinuities, or in the context of Mendellian randomization, these causal effects and estimators may apply. To facilitate implementation, we provide an easy-to-use R package and step-by-step instructions \url{https://github.com/nt-williams/iv_mediation}.

In numerical studies, the performance of our proposed estimators deteriorated in small samples, e.g., N=500. This is not surprising. Estimators of mediational direct and indirect effects are more sensitive to smaller sample sizes than total effect estimators \citep{rudolph2020peril}. Estimators of complier total effects (e.g., complier average causal effects) are more sensitive to smaller sample sizes that 
average treatment effect estimators \citep{jo2002statistical}. Given that we consider \textit{both} mediational \textit{and} complier estimators here, we would expect finite sample performance to be challenged even further. It reassuring, however, that performance was acceptable in the N=1,000 sample size tested in our simulations.

Future work could focus on making several improvements to our estimators, including: developing an estimator in the targeted minimum loss-based framework that targets each ratio directly instead separately estimating the numerator and denominator, and incorporating the monotonicity assumption into the estimation process.

\newpage
\begin{appendix}
\large{\textbf{Supplementary Materials for \titlepaper}}
\vspace{1cm}
\renewcommand{\thesection}{S\arabic{section}}
\renewcommand{\thesubsection}{S\arabic{subsection}}
\section{Identification proofs}
\subsection{Identification of the complier interventional direct effect}
\begin{footnotesize}
\begin{proof}

\begin{align*}
\E(Y_{Z=1,M=G_{Z=0}} - Y_{Z=0,M=G_{Z=0}} \mid C_Z=1)\\
& \text{using the definition of } C_Z=1, \text{so } Z_{A=a}=a  \implies\\
&= \E(Y_{Z=Z_1,M=G_{Z=Z_0}} - Y_{Z=Z_0,M=G_{Z=Z_0}} \mid C_Z=1)\\
& \text{by the law of iterated expectation} \implies \\
&= \int_{\mathcal{W}} \E(Y_{Z=Z_1,M=G_{Z=Z_0}} - Y_{Z=Z_0,M=G_{Z=Z_0}} \mid C_Z=1, w) \dd\P(w \mid C_Z=1)\\
\end{align*}

Note that:
\begin{align*}
 \E(Y_{Z_1,G_{Z_0}} - Y_{Z_0,G_{Z_0}} \mid w) &= \E(Y_{Z_1,G_{Z_0}} - Y_{Z_0,G_{Z_0}} \mid C_Z=1, w) \P(C_Z=1 \mid w)\\
&+  \E(Y_{Z_1,G_{Z_0}} - Y_{Z_0,G_{Z_0}} \mid C_Z=0, w) \P(C_Z=0 \mid w)\\
&+  \E(Y_{Z_1,G_{Z_0}} - Y_{Z_0,G_{Z_0}} \mid C_Z= -1, w) \P(C_Z= -1 \mid w)\\
& \ref{ass:mono} \implies \\
&=  \E(Y_{Z_1,G_{Z_0}} - Y_{Z_0,G_{Z_0}m} \mid C_Z=1, w) \P(C_Z=1 \mid w)\\
&+  \E(Y_{Z_1,G_{Z_0}} - Y_{Z_0,G_{Z_0}} \mid C_Z=0, w) \P(C_Z=0 \mid w)\\
& \text{when $C_Z=0$ then } Z_{1} = Z_0 \implies\\
&=  \E(Y_{Z_1,G_{Z_0}} - Y_{Z_0,G_{Z_0}} \mid C_Z=1, w) \P(C_Z=1 \mid w)
\end{align*}

Using the above result, we can continue the proof from above:
\begin{align*}
&= \int_{\mathcal{W}} \frac{\E(Y_{Z_1,G_{Z_0}} - Y_{Z_0,G_{Z_0}} \mid w)}{\P(C_Z=1 \mid w)} \dd\P(w \mid C_Z=1)\\
&= \int_{\mathcal{W}} \frac{\E(Y_{Z_1,G_{Z_0}} - Y_{Z_0,G_{Z_0}} \mid w)}{\P(C_Z=1 \mid w)} \frac{\P(C_Z =1 \mid w)}{\P(C_Z=1)}\dd\P(w)\\
&= \int_{\mathcal{W}} \frac{\E(Y_{Z_1,G_{Z_0}} - Y_{Z_0,G_{Z_0}} \mid w)\dd\P(w)}{\P(C_Z=1)} \\
& \ref{ass:exclr} \implies \\
&= \int_{\mathcal{W}} \frac{\E(Y_{A=1,G_{A=0}} - Y_{A=0,G_{A=0}} \mid w)\dd\P(w)}{\P(C_Z=1)}\\
&= \int_{\mathcal{W}, \mathcal{M}} \frac{\E(Y_{A=1,M=m} - Y_{A=0,M=m} \mid w) \dd\P(G_{A_0}=m | w)\dd\P(w)}{\P(C_Z=1)} \\
&= \int_{\mathcal{W}, \mathcal{M}} \frac{\E(Y_{A=1,M=m} - Y_{A=0,M=m} \mid w) \dd\P(M_{A=0}=m | w)\dd\P(w)}{\P(C_Z=1)} \\
&\text{using the result from above}\\
&=  \int_{\mathcal{W}, \mathcal{M}}\frac{ \E(Y_{A=1,M=m} - Y_{A=0,M=m} \mid w) \dd\P(M_{A=0}=m | w)}{\E(Z_{A=1} - Z_{A=0}\mid w)} \dd\P(w)\\
&\ref{ass:exch} \implies\\
&= \int_{\mathcal{W}, \mathcal{Z}, \mathcal{M}} \{[ (\E(Y_{A=1,M=m} \mid A=1,z,  m, w)\P(Z=z \mid A=1, w) \\ 
&- \E(Y_{A=0,M=m} \mid A=0, z, m, w)\P(Z=z \mid A=0, w)) \dd\P(M_{A=0}=m | A=0, w) ] \\
&/ [\E(Z_{A=1} \mid A=1, w) - \E(Z_{A=0}\mid A=0, w) ]\} \dd\P(w)\\
& \text{consistency} \implies\\
&= \int_{\mathcal{W}, \mathcal{Z}, \mathcal{M}} \frac{[\E(Y\mid A=1,z,  m, w)\P(z \mid A=1, w) - \E(Y \mid A=0, z, m, w)\P(z \mid A=0, w)] \dd\P(m | A=0, w) }{ \E(Z\mid A=1, w) - \E(Z\mid A=0, w) }\dd\P(w) \\
& \text{by assumptions \ref{ass:pos} and \ref{ass:nonzeroeffect} we have that the above is defined.}
\end{align*}
\end{proof}
\end{footnotesize}

\subsection{Identification of the double complier interventional indirect effect}
\begin{footnotesize}
\begin{proof}

\begin{align*}
\E&(Y_{Z=1,M=G_{Z=1}} - Y_{Z=1,M=G_{Z=0}} \mid C_Z=1, C_M=1)\\
&\text{using the definition of } C_Z=1, \text{so } Z_{A=a}=a  \implies\\
&= \E(Y_{Z=Z_1,M=G_{Z=Z_1}} - Y_{Z=Z_1,M=G_{Z=Z_0}} \mid C_Z=1, C_M=1)\\
& \text{by the law of iterated expectation} \implies \\
&= \int_{\mathcal{W}} \E(Y_{Z=Z_1,M=G_{Z=Z_1}} - Y_{Z=Z_1,M=G_{Z=Z_0}} \mid C_Z=1, C_M=1, w) \dd\P(w \mid C_Z=1, C_M=1)\\
&= \int_{\mathcal{W}, \mathcal{M}_1,\mathcal{M}_2} [\E(Y_{Z=Z_1,M=m_1} \mid C_Z=1, C_M=1, w, G_{Z_1}=m_1) \dd\P(G_{Z_1}=m_1 \mid C_Z=1, C_M=1, w) \\
&- \E(Y_{Z=Z_1,M=m_2} \mid C_Z=1, C_M=1, w, G_{Z_0}=m_2)\dd\P(G_{Z_0}=m_2 \mid C_Z=1, C_M=1, w)  ] \dd\P(w \mid C_Z=1, C_M=1)\\
&= \int_{\mathcal{W}, \mathcal{M}_1,\mathcal{M}_2} [\E(Y_{Z=Z_1,M=m_1} \mid C_Z=1, C_M=1, w) \dd\P(M_{Z_1}=m_1 \mid w) \\
&- \E(Y_{Z=Z_1,M=m_2} \mid C_Z=1, C_M=1, w)\dd\P(M_{Z_0}=m_2 \mid w)   ] \dd\P(w \mid C_Z=1, C_M=1)\\
&= \int_{\mathcal{W}, \mathcal{M}_1,\mathcal{M}_2} \E(Y_{Z=Z_1,M=m_1} - Y_{Z=Z_1,M=m_2} \mid C_Z=1, C_M=1, w) \\
&\dd\P(M_{Z_1}=m_1 \mid w)\dd\P(M_{Z_0}=m_2 \mid w) \dd\P(w \mid C_Z=1, C_M=1)\\
\end{align*}

Assume $C_Z=C_M$ (as in \ref{ass:mono2}) and note that:
\begin{align*}
 &\E(Y_{Z=Z_1,L=m1} - Y_{Z=Z_1,L=m2} \mid w) \\
 & \ref{ass:exclr2} \implies \\
 &= \E(Y_{Z=Z_1,M=M_{L=m1}} - Y_{Z=Z_1,M=M_{L=m2}} \mid w, C_Z=C_M=1)  \P(C_Z=C_M=1 \mid w) \\
 &+\E(Y_{Z=Z_1,M=M_{L=m1}} - Y_{Z=Z_1,M=M_{L=m2}} \mid w, C_Z=C_M=0)  \P(C_Z=C_M=0 \mid w) \\
 &+\E(Y_{Z=Z_1,M=M_{L=m1}} - Y_{Z=Z_1,M=M_{L=m2}} \mid w, C_Z=C_M= -1)  \P(C_Z=C_M= -1 \mid w) \\
& \ref{ass:mono2} \implies\\
&= \E(Y_{Z=Z_1,M=M_{L=m1}} - Y_{Z=Z_1,M=M_{L=m2}} \mid w, C_Z=C_M=1)  \P(C_Z=C_M=1 \mid w) \\
 &+\E(Y_{Z=Z_1,M=M_{L=m1}} - Y_{Z=Z_1,M=M_{L=m2}} \mid w, C_Z=C_M=0)  \P(C_Z=C_M=0 \mid w) \\
 & \text{when $C_M=0$ then } M_{L=m1} = M_{L=m2}  \implies\\
&= \E(Y_{Z=Z_1,M=M_{L=m1}} - Y_{Z=Z_1,M=M_{L=m2}} \mid w, C_Z=C_M=1)  \P(C_Z=C_M=1 \mid w) \\
\end{align*}

Using this result, we can continue the proof from above:
\begin{align*}
&= \int_{\mathcal{W}, \mathcal{M}_1,\mathcal{M}_2} \frac{\E(Y_{Z=Z_1,L=m_1} - Y_{Z=Z_1,L=m_2} \mid w) \dd\P(M_{Z_1}=m_1 \mid w)\dd\P(M_{Z_0}=m_2 \mid w) }{\P(C_Z=1, C_M=1 \mid w)}\dd\P(w \mid C_Z=1, C_M=1)\\
&= \int_{\mathcal{W}, \mathcal{M}_1,\mathcal{M}_2} \frac{\E(Y_{Z=Z_1,L=m_1} - Y_{Z=Z_1,L=m_2} \mid w) \dd\P(M_{Z_1}=m_1 \mid w)\dd\P(M_{Z_0}=m_2 \mid w) }{\P(C_Z=1, C_M=1 \mid w)}\frac{\P(C_Z=1, C_M=1 \mid w)\dd\P(w)}{\P(C_Z=1, C_M=1)}\\
&= \int_{\mathcal{W}, \mathcal{M}_1,\mathcal{M}_2} \frac{\E(Y_{Z=Z_1,L=m_1} - Y_{Z=Z_1,L=m_2} \mid w) \dd\P(M_{Z_1}=m_1 \mid w)\dd\P(M_{Z_0}=m_2 \mid w) \dd\P(w)}{\P(C_Z=1, C_M=1)}\\
& \ref{ass:exclr2} \implies \\
&= \int_{\mathcal{W}, \mathcal{M}_1,\mathcal{M}_2} \frac{\E(Y_{A=1,L=m_1} - Y_{A=1,L=m_2} \mid w) \dd\P(M_{A=1}=m_1 \mid w)\dd\P(M_{A=0}=m_2 \mid w) \dd\P(w)}{\P(C_Z=1, C_M=1)}\\
&= \int_{\mathcal{W}, \mathcal{M}_1,\mathcal{M}_2} \frac{\E(Y_{A=1,L=m_1} - Y_{A=1,L=m_2} \mid w) \dd\P(M_{A=1}=m_1 \mid w)\dd\P(M_{A=0}=m_2 \mid w) \dd\P(w)}{\E(C_ZC_M)}\\
&= \int_{\mathcal{W}, \mathcal{M}_1,\mathcal{M}_2} \frac{\E(Y_{A=1,L=m_1} - Y_{A=1,L=m_2} \mid w) \dd\P(M_{A=1}=m_1 \mid w)\dd\P(M_{A=0}=m_2 \mid w) \dd\P(w)}{\E[(M_{L=1} - M_{L=0}) (Z_{A=1}-Z_{A=0})]}\\
&= \int_{\mathcal{W}, \mathcal{M}_1,\mathcal{M}_2} \frac{\E(Y_{A=1,L=m_1} - Y_{A=1,L=m_2} \mid w) \dd\P(M_{A=1}=m_1 \mid w)\dd\P(M_{A=0}=m_2 \mid w) \dd\P(w)}{\E(M_{L=1}Z_{A=1} - M_{L=0}Z_{A=1} - M_{L=1}Z_{A=0} + M_{L=0}Z_{A=0} )}
\end{align*}
The numerator can be identified as follows:
\begin{align*}
&\int_{\mathcal{W}, \mathcal{M}_1,\mathcal{M}_2} \E(Y_{A=1,L=m_1} - Y_{A=1,L=m_2} \mid w) \dd\P(M_{A=1}=m_1 \mid w)\dd\P(M_{A=0}=m_2 \mid w) \dd\P(w)\\
&\ref{ass:exch2} \implies \\
&= \int_{\mathcal{W}, \mathcal{Z}, \mathcal{M}_1,\mathcal{M}_2} [\E(Y_{A=1,L=m_1} \mid A=1, z, L=m_1, w)\P(z \mid A=1, w)\dd\P(M_{A=1}=m_1 \mid A=1, w) \\
&- \E(Y_{A=1,L=m_2} \mid w, A=1,z, L=m_2) \P(z \mid A=1, w)\dd\P(M_{A=0}=m_2 \mid A=0, w)] \dd\P(w)\\
&\text{consistency} \implies \\
&= \int_{\mathcal{W}, \mathcal{Z}, \mathcal{M}} \E(Y\mid A=1, z, L=m, w)\P(z \mid A=1, w)[\P(M=m \mid A=1, w) - \P(M=m \mid A=0, w)] \dd\P(w)
\end{align*}
The denominator can be identified following the identification for a two-timepoint longitudinal intervention:
\begin{align*}
& \E[M_{L=1}Z_{A=1} - M_{L=0}Z_{A=1} - M_{L=1}Z_{A=0} + M_{L=0}Z_{A=0} ]\\
&\text{define a new outcome, $Q=MZ$, and counterfactual outcome} Q_{l,a}=M_{L=l}Z_{A=a}\\ 
&= \E(Q_{1,1} - Q_{0,1} - Q_{1,0} + Q_{0,0})\\
&\ref{ass:exch2} \implies \\
&= \int_{\mathcal{W}}[\E(Q_{1,1} \mid A=1, w) - \E(Q_{0,1} \mid A=1, w) - \E(Q_{1,0} \mid A=0, w) + \E(Q_{0,0} \mid A=0, w)] \dd\P(w)\\
&\ref{ass:exch2} \implies \\
&= \int_{\mathcal{W}, \mathcal{Z}}[\E(Q_{1,1} \mid L=1, z, A=1, w)\dd\P(z \mid A=1, w) - \E(Q_{0,1} \mid L=0, z, A=1, w)\dd\P(z \mid A=1, w)\\
&- \E(Q_{1,0} \mid L=1, z, A=0, w)\dd\P(z \mid A=0, w) + \E(Q_{0,0} \mid L=0, z, A=0, w)\dd\P(z \mid A=0, w)] \dd\P(w)\\
&\text{consistency} \implies \\
&= \int_{\mathcal{W}, \mathcal{Z}}[\E(MZ \mid L=1, z, A=1, w)\dd\P(z \mid A=1, w) - \E(MZ \mid L=0, z, A=1, w)\dd\P(z \mid A=1, w)\\
&- \E(MZ \mid L=1, z, A=0, w)\dd\P(z \mid A=0, w) + \E(MZ \mid L=0, z, A=0, w)\dd\P(z \mid A=0, w)] \dd\P(w)\\
&= \int_{\mathcal{W}, \mathcal{Z}} [z \P(M=1 \mid L=1, z, w)\dd\P(z \mid A=1, w) - z \P(M=1 \mid L=0, z, w)\dd\P(z \mid A=1, w) \\
&- z \P(M=1\mid L=1, z, w)\dd\P(z \mid A=0, w) + z \P(M=1 \mid L=0, z,w)\dd\P(z \mid A=0, w) ] \dd\P(w)\\
&= \int_{\mathcal{W}} [\P(M=1 \mid L=1, Z=1, w)\P(Z=1 \mid A=1, w) -  \P(M=1 \mid L=0, Z=1, w)\P(Z=1 \mid A=1, w) \\
&- \P(M=1\mid L=1, Z=1, w)\P(Z=1 \mid A=0, w) + \P(M=1 \mid L=0, Z=1,w)\P(Z=1 \mid A=0, w) ] \dd\P(w)\\
&\text{by assumptions \ref{ass:nonzeroeffect2} and \ref{ass:pos2} we have that the above is defined}
\end{align*}
\end{proof}
\end{footnotesize}

\subsection{Identification of the double complier interventional direct effect}
\begin{footnotesize}
\begin{proof}

\begin{align*}
&\E(Y_{Z=1,M=G_{Z=0}} - Y_{Z=0,M=G_{Z=0}} \mid C_Z=1)\\
& \text{using the definition of } C_Z=1, \text{so } Z_{A=a}=a  \implies\\
&= \E(Y_{Z=Z_1,M=G_{Z=Z_0}} - Y_{Z=Z_0,M=G_{Z=Z_0}} \mid C_Z=1)\\
& \text{by the law of iterated expectation} \implies \\
&= \int_{\mathcal{W}} \E(Y_{Z=Z_1,M=G_{Z=Z_0}} - Y_{Z=Z_0,M=G_{Z=Z_0}} \mid C_Z=1, w) \dd\P(w \mid C_Z=1)\\
&= \int_{\mathcal{W}, \mathcal{M}} \E(Y_{Z=Z_1,M=m} - Y_{Z=Z_0,M=m} \mid C_Z=1, w, G_{A_0}=m) \dd\P(G_{A_0}=m | C_Z=1, w) \dd\P(w \mid C_Z=1)\\
&= \int_{\mathcal{W}, \mathcal{M}} \E(Y_{Z=Z_1,M=m} - Y_{Z=Z_0,M=m} \mid C_Z=1, w) \dd\P(M_{Z=Z_0}=m | w) \dd\P(w \mid C_Z=1)
\end{align*}

Note that:
\begin{align*}
 \E(Y_{Z_1,L=m} - Y_{Z_0,L=m} \mid w) \\
 &\ref{ass:exch2} \implies\\
 &= \E(Y_{Z_1,M=M_{L=m}} - Y_{Z_0,M=M_{L=m}} \mid C_Z=1, w) \P(C_Z=1 \mid w)\\
&+  \E(Y_{Z_1,M_{L=m}} - Y_{Z_0,M_{L=m}} \mid C_Z=0, w) \P(C_Z=0 \mid w)\\
&+  \E(Y_{Z_1,M_{L=m}} - Y_{Z_0,M_{L=m}} \mid C_Z= -1, w) \P(C_Z= -1 \mid w)\\
& \ref{ass:mono} \implies \\
&=  \E(Y_{Z_1,M_{L=m}} - Y_{Z_0,M_{L=m}} \mid C_Z=1, w) \P(C_Z=1 \mid w)\\
&+  \E(Y_{Z_1,M_{L=m}} - Y_{Z_0,M_{L=m}} \mid C_Z=0, w) \P(C_Z=0 \mid w)\\
& \text{when $C_Z=0$ then } Z_{1} = Z_0 \implies\\
&=  \E(Y_{Z_1,M_{L=m}} - Y_{Z_0,M_{L=m}} \mid C_Z=1, w) \P(C_Z=1 \mid w)
\end{align*}

Using the above result, we can continue the proof from above:
\begin{align*}
&= \int_{\mathcal{W}, \mathcal{M}} \frac{\E(Y_{Z_1,L=m} - Y_{Z_0,L=m} \mid w)\dd\P(M_{Z_0}=m | w)}{\P(C_Z=1 \mid w)} \dd\P(w \mid C_Z=1)\\
&= \int_{\mathcal{W}, \mathcal{M}} \frac{\E(Y_{Z_1,L=m} - Y_{Z_0,L=m} \mid w)\dd\P(M_{Z_0}=m | w)}{\P(C_Z=1 \mid w)} \frac{\P(C_Z =1 \mid w)}{\P(C_Z=1)}\dd\P(w)\\
&= \int_{\mathcal{W}, \mathcal{M}} \frac{\E(Y_{Z_1,L=m} - Y_{Z_0,L=m} \mid w)\dd\P(M_{Z_0}=m | w)\dd\P(w)}{\P(C_Z=1)} \\
& \ref{ass:exclr} \implies \\
&= \int_{\mathcal{W}, \mathcal{M}} \frac{\E(Y_{A=1,L=m} - Y_{A=0,L=m} \mid w)\dd\P(M_{A=0}=m | w)\dd\P(w)}{\P(C_Z=1)}\\
&\text{using the result from above}\\
&=  \int_{\mathcal{W}, \mathcal{M}}\frac{ \E(Y_{A=1,L=m} - Y_{A=0,L=m} \mid w) \dd\P(M_{A=0}=m | w)}{\E(Z_{A=1} - Z_{A=0}\mid w)} \dd\P(w)
\end{align*}
The numerator can be identified as follows:
\begin{align*}
&\int_{\mathcal{W}, \mathcal{M}} \E(Y_{A=1,L=m} - Y_{A=0,L=m} \mid w) \dd\P(M_{A=0}=m \mid w) \dd\P(w)\\
&\ref{ass:exch3} \implies \\
&= \int_{\mathcal{W}, \mathcal{Z}, \mathcal{M}} [\E(Y_{A=1,L=m} \mid A=1, z, L=m, w)\dd\P(z \mid A=1, w)\\
&- \E(Y_{A=0,L=m} \mid w, A=0,z, L=m) \dd\P(z \mid A=0, w)]\dd\P(M_{A=0}=m \mid A=0, w) \dd\P(w)\\
&\text{consistency} \implies \\
&= \int_{\mathcal{W}, \mathcal{Z}, \mathcal{M}} [\E(Y\mid A=1, z, L=m, w)\dd\P(z \mid A=1, w)\\
&- \E(Y \mid w, A=0,z, L=m) \dd\P(z \mid A=0, w)]\dd\P(M_{A=0}=m \mid A=0, w) \dd\P(w)\\
\end{align*}
The denominator can be identified as follows:
\begin{align*}
 &\int_{\mathcal{W}} \E(Z_{A=1} - Z_{A=0}\mid w)\dd\P(w)\\
 &\ref{ass:exch3} \implies\\
&\int_{\mathcal{W}} \E(Z_{A=1} \mid A=1, w) - \E(Z_{A=0}\mid A=0, w) \dd\P(w)\\
& \text{consistency} \implies\\
&\int_{\mathcal{W}} \E(Z\mid A=1, w) - \E(Z\mid A=0, w) \dd\P(w) \\
& \text{by assumptions \ref{ass:pos2} and \ref{ass:nonzeroeffect} we have that the above is defined.}
\end{align*}
\end{proof}
\end{footnotesize}

\section{Efficient influence functions derived by others}
The EIF for $\psi_{FS}$ was derived in \citet{hahn1998role}.
\begin{equation}
    D_{FS}(O, \P) = \frac{2A-1}{\g(a | W)}(Z-\q(1 | a,W)) + \q(1,W) - \q(0,W)- \psi_{FS}.
\end{equation} 

The EIF for $\psi_{IDE}$ was derived in \citet{diaz2019non}.

\begin{align}
\begin{split}
D_{IDE}(O, \P) = & D^{a^\prime=1, a^\star=0}(O, \P) - D^{a^\prime=0, a^\star=0}(O, \P), 
  \end{split}
\end{align} where 
\begin{align}
\begin{split}
D^{a^\prime, a^\star}(O, \P) = &\frac{\one\{A=a^{\prime}\}}{\g(a^{\prime}\mid W)}
                    \h(Z, M, W)\{Y - \mu(Z, M, W)\}
  \\
  + & \frac{\one\{A=a^{\prime}\}}{\g(a^{\prime}\mid W)}\left\{\uu(Z,W)-\int_{\mathcal
            Z}\uu(z,W)\q(z\mid a^{\prime},W)\dd\nu(z)\right\}
            \\
  +  &  \frac{\one\{A=a^{\star}\}}{\g(a^{\star}\mid W)}\left\{\int_{\mathcal
            Z}\mu(a^{\prime},z,M,W)\q(z\mid a^{\prime},W)\dd\nu(z)-\vv(W) \right\}\\
            + & \vv(W) -\theta(a', a^\star),
  \end{split}
\end{align}

where

\begin{align}
        \h(Z, M, W) = &  \frac{\q(Z\mid a^{'},W)}{\rr(Z\mid a^{'},M, W)} \frac{\e(a^{\star}\mid M, W)}{\e(a^{'}\mid M, W)}\\
        \uu(z, W) = & \E\left\{\mu(z,M,W)\h(z,M,W),\bigg|\,
                 z,a^{\prime},W\right\},\\
        \vv(W) = & \E\left\{\int_{\mathcal
               Z}\mu(z,M,W)\q(z\mid a^{\prime},W)\dd\nu(z)\,\bigg|\, a^\star,W\right\}.
\end{align}

\section{Baseline covariates used in the empirical illustration}
 \begin{itemize}
     \item Adolescent characteristics: site (Boston, Chicago, LA, NYC), age, race/ethnicity (categorized as black, latino/Hispanic, white, other), number of family members (categorized as 2, 3, or 4+), someone from school asked to discuss problems the child had with schoolwork or behavior during the 2 years prior to baseline, child enrolled in special class for gifted and talented students.
     \item Adult household head characteristics included: high school graduate, marital status (never vs ever married), whether had been a teen parent, work status, receipt of AFDC/TANF, whether any family member has a disability.
     \item Neighborhood characteristics: felt neighborhood streets were unsafe at night; very dissatisfied with neighborhood; poverty level of neighborhood.
     \item Reported reasons for participating in MTO: to have access to better schools.
     \item Moving-related characteristics: moved more then 3 times during the 5 years prior to baseline, previous application for Section 8 voucher.
 \end{itemize}
\end{appendix}
\bibliographystyle{plainnat}
\bibliography{refs}
\end{document}